\documentclass[12pt,preprint]{aastex}

\def\spose#1{\hbox to 0pt{#1\hss}}

\def\lsim{\mathrel{\hbox{\rlap{\lower.55ex \hbox {$\sim$}}\kern-.0em\raise.4ex \hbox{$<$}}}} 
\def\gsim{\mathrel{\hbox{\rlap{\lower.55ex \hbox {$\sim$}}\kern-.0em\raise.4ex \hbox{$>$}}}}

\begin{document}

\title{A Study of the Afterglows of Four GRBs: Constraining the
Explosion and Fireball Model}

\author{S. A. Yost\altaffilmark{1},
 F. A. Harrison\altaffilmark{1},
 R.    Sari\altaffilmark{2},
 D. A. Frail\altaffilmark{3},
}

\begin{abstract}

We employ a fireball model of the gamma-ray burst explosion to
constrain intrinsic and environmental parameters of four events with
good broadband afterglow data; GRB~970508, GRB~980329, GRB~980703, and
GRB~000926.  Using the standard assumptions of constant circumburst
density and no evolution of the fraction of the explosion energy in
the post-shock magnetic field, we investigate the uniformity of the
derived explosion and shock physics parameters among these events. We
find a variety of parameters: densities that range from those of the
ISM to diffuse clouds, energies comparable to the total GRB
$\gamma$-ray energy, collimations from near-isotropy to 0.04 radians,
substantial electron energy fractions of 10-30\% with energy
distribution indices of 2.1-2.9 and magnetic energy fractions from
0.2-25\%.  We also investigate the level to which the data constrain
the standard model assumptions, such as the magnetic field evolution,
and the allowed density profiles of the medium. Fits generally improve
slightly with an increasing magnetic energy fraction
$\epsilon_B$. Good fits can be produced with magnetic energy
accumulating or decaying with the shock strength over the afterglow as
$\epsilon_B \propto \gamma^{x}$; $-2 \leq x \leq +1$ . The data are
not very sensitive to increasing density profiles, allowing good fits
even with density $\propto r^{10}$. Some parameter values change by up
to an order of magnitude under such altered assumptions; the
parameters of even good fits cannot be taken at face value. The data
are sensitive to decreasing densities; $r^{-2}$ profiles may produce
reasonable fits, but steeper profiles, even $r^{-2.5}$, will not fit
the data.

\end{abstract}

\keywords{gamma rays: bursts}

\slugcomment{Submitted to The Astrophysical Journal}
\altaffiltext{1}{Division of Physics, Mathematics and Astronomy,
  105-24, California Institute of Technology, Pasadena, CA 91125}
\altaffiltext{2}{Theoretical Astrophysics 130-33, California Institute
  of Technology, Pasadena, CA 91125}
\altaffiltext{3}{National Radio Astronomy Observatory, P.O. BOX `O', Socorro,
NM 87801}

\section{Introduction}

Afterglow emission, spanning from radio to X-ray, and observable from
minutes to weeks after the event, appears to be a ubiquitous feature
of long-duration gamma-ray burst explosions.  Afterglows result from
radiation emitted by electrons accelerated in a relativistic shock
produced by the explosion of a progenitor \citep[e.g.,][]{pr93, k94}.
As the shock propagates outward into the surrounding medium, the
resulting broadband synchrotron radiation evolves in a manner
dependent on a number of fundamental characteristics of the explosion, as
well as on the details of the shock evolution and the density profile of the
medium it expands into \citep[for a review of the theory, see][]{m02}.

The physics of relativistic shocks, the mechanism of collisionless
shock acceleration, and the means by which magnetic fields can be
amplified to levels needed to produce the observed synchrotron
emission are poorly understood.  Basic understanding of these elements
will eventually result from improved magnetohydrodynamic simulations,
which may ultimately prescribe such factors as the appropriate
electron energy distribution, what fraction of shock energy goes into
magnetic field, and how this evolves with the shock expansion.
Additional uncertainties arise from the fact that the outflow
geometry, and the structure of the circumburst medium are
unobservable, and could potentially be complex.

In spite of the lack of detailed understanding, the basic features of
the GRB afterglow can be described by relatively simple theoretical
models for the outflow and shock \citep[e.g. with relativistic flow as
in][]{bm76}. Simple assumptions about the shock microphysics and
geometry appear to fit the basic features of the afterglow in a number
of cases
\citep[e.g.,][]{wg99, cl00, hys+01, pk01, pk01b, pk02}.  In addition,
many events appear consistent with the expansion of the shock into a
medium with either a constant or a simple $r^{-2}$ powerlaw in density.

Ideally, high-quality broadband afterglow observations, interpreted in
the context of basic theoretical models, could be used to constrain
the explosion parameters, geometry, and the structure of the
surrounding medium.  In addition, if the data are of sufficiently
high quality, it should be possible to test the validity of the basic
model assumptions; or at least the range over which they may be varied
and still describe the data.  Given a relatively large number of
model parameters, however, only a few data sets are of sufficient
quality to provide interesting constraints.

In this paper, we examine four high-quality broadband afterglow data
sets, fitting them to a basic afterglow model.  We investigate the
similarity of the derived explosion, shock, and environmental
parameters.  In addition, we investigate selected model assumptions
and the range over which they can be varied.  In particular, we
consider whether the fraction of energy in the post-shock magnetic
field can evolve as some power of the bulk Lorentz factor, and the
range of possible matter density gradients.  Finally, we consider
which future observations provide the most promise for better
constraining both the model and its physical parameters.

\section{The Afterglow Framework}\label{sec:framework}

We adopt a standard fireball scenario for the GRB afterglow, where a
relativistic shock with Lorentz factor $\gamma$ expands into the
circumburst medium (CBM). \citet{bm76} describes the flow. The shock
may be spherical or initially confined to a cone (i.e., be
jet-like). The afterglow flux arises from the radiation (synchrotron
and possibly also inverse Compton) emitted by relativistic electrons
accelerated in the shock.  To describe its evolution we model the
shock dynamics, adopting assumptions for the density profile of the
CBM, the shock geometry, and the shock microphysics including the
distribution of electron energies, as described below. We also
account for the effects of the medium through which the radiation
passes en route to the observer.

\subsection{The Basic Afterglow Model}\label{sec:basicmodel}

For the dynamics, we make simple assumptions about the event's
geometry and environment. We consider a constant circumburst matter
density $n$, the ``ISM-like'' case, as well as the possibility the CBM
could be dominated by a wind outflow from the progenitor. As a
constant mass-loss rate and wind speed gives an $r^{-2}$ profile, we
consider this the ``Wind-like'' case. Into either CBM form we allow
for isotropy or a simple collimation of the ejecta, a top-hat
distribution in solid angle with half-opening angle $\theta$. The
shock behaves as though isotropic until it slows down sufficiently to
expand in its rest frame \citep{rho99,sph99}. We calculate the time
$t_{jet}$ at which collimation is evident as $\theta$ = 1/$\gamma$
($\gamma$ the shock Lorentz factor; this assumes the observer is
nearly along the line of sight). We calculate the expected radiative
losses, which can modify the shock dynamics \citep{sar97}, in the
manner detailed further below.

For the microphysics governing the emission it is standard to assume
several things. First, that the shock imparts a constant fraction of
its energy ($\epsilon_e$) to the swept-up electrons, and a constant
fraction ($\epsilon_B$) goes into amplifying magnetic fields. These
fractions are capped at 100\%, where all the shock's energy would be
in one of these. A reasonable limit could have been the
ejecta-$\epsilon_e$-$\epsilon_B$ equipartition value of 33\%, but the
model's calculations are not perfectly known. The equations used have
some uncertainties; we allow each of $\epsilon_e$ and $\epsilon_B$
to vary up to the limit of taking all the shock energy. It is also
assumed that the electrons are accelerated into a simple powerlaw
distribution of energies above a minimum value (P($\gamma_e$)$\propto
\gamma_e^{-p}$, $\gamma_e > \gamma_m$), with a constant index $p$.

These are our basic assumptions for the fireball model. They will be
tested in further sections.

\subsection{The Emission calculation}\label{sec:emission}

Given this electron spectrum, we get a broken powerlaw radiation
spectrum, with three break frequencies that evolve in a manner that
depends upon the dynamics \citep{spn98}: the injection break $\nu_m$
for the minimal energy of the radiating electrons, the cooling break
$\nu_c$ corresponding to energies at which radiative losses over the
shock's lifetime are significant and the self-absorption break $\nu_a$
where the spectrum becomes optically thick at low frequencies. If the
minimum energy electrons emit at a peak frequency above the
self-absorbed regime, the spectrum below $\nu_a$ is $\propto \nu^2$,
ressembling a blackbody with effective temperature corresponding to
that minimum energy; if electrons are emitting within the optically
thick regime, the effective temperature is a function of frequency and
the spectrum is $\propto \nu^{5/2}$ \citep{rl79}. We use a smooth
shape with factors of (1 + $(\nu/\nu_{break})^{\beta_1 -
\beta_2}$)$^{-1}$ ($\beta_{1,2}$ are the indices before and after the
break) for $\nu_m$ and $\nu_c$, and for $\nu_a$ we use the physically
motivated prescription of \citet{gps99b}.

For the model with basic assumptions we employ the previously
published fireball model calculations for the spectral breaks. The
isotropic case is taken from \citet{spn98}'s relativistic
near-adiabatic case, with normalizations from \citet{gs02} (their
canonical order, Table 2) for a Wind-like CBM, and from
\citet{spn98,gps99b,gps99c} for an ISM-like CBM. These normalizations
account for the post-shock spatial distribution of electrons and
emission from equal arrival-time surfaces. The resulting equations are:

\begin{equation}
\begin{array}{lccc}
F_{\nu}^{max} & = & 1.6 \, (z+1) \,D_{28}^{-2}\, \epsilon_{B,-2}^{0.5}\,
E_{52} \,n^{0.5} & {\rm mJy}\\
\nu_a & = & 4.2\times10^{8}  \, (z+1)^{-1}\, f_I(p)\, \bar{\epsilon}_e^{\,-1}\, \epsilon_{B,-2}^{0.2}\, E_{52}^{0.2}\, n^{0.6}  & {\rm Hz}\\
\nu_m & = & 3.3\times10^{14} \,  (z+1)^{0.5}\, \epsilon_{B,-2}^{0.5} \,\bar{\epsilon}_e^{2}\, E_{52}^{0.5}\, t_d^{-1.5} & {\rm Hz}\\
\nu_c & = & 6.3\times10^{15}  \, (z+1)^{-0.5}\, \epsilon_{B,-2}^{-1.5} \,E_{52}^{-0.5} \,n^{-1} \,t_d^{-0.5} & {\rm Hz}\\
\end{array}
\end{equation}
For the ISM-like CBM, where $f_I(p) = (\,(p + 2)(p - 1)/(3p + 2)\,)^{0.6}$ and
$\bar{\epsilon}_e = \epsilon_e(p-2)/(p-1)$. For the Wind-like CBM:
\begin{equation}
\begin{array}{lccc}
F_{\nu}^{max} & = &  7.7\, (p+0.12)\, (z+1)^{1.5} \,D_{28}^{-2}\, \epsilon_{B,-2}^{0.5}\,
E_{52} \, A_{*} \, t_d^{-0.5} & {\rm mJy}\\
\nu_a & = & 3.3\times10^{9}  \, (z+1)^{-0.4}\, f_W(p)\, \bar{\epsilon}_e^{\,-1}\, \epsilon_{B,-2}^{0.2}\, E_{52}^{-0.4}\, A_{*}^{1.2} \, t_d^{-0.6} & {\rm Hz}\\
\nu_m & = & 4.0\times10^{14} \, (p-0.69)\, (z+1)^{0.5}\, \epsilon_{B,-2}^{0.5} \,\bar{\epsilon}_e^{2}\, E_{52}^{0.5}\, t_d^{-1.5} & {\rm Hz}\\
\nu_c & = & 4.4\times10^{13} \, (3.45-p)e^{0.45p} \, (z+1)^{-1.5}\, \epsilon_{B,-2}^{-1.5} \,E_{52}^{0.5} \,A_{*}^{-2} \,t_d^{0.5} & {\rm Hz}\\
\end{array}
\end{equation}
where $f_W(p) = (\,(p - 1)/(3p + 2)\,)^{0.6}$. The units are: $D_{28}$
= luminosity distance in 10$^{28}$~cm, $\epsilon_{B,-2}$ =
$\epsilon_B$ in \%, $n$ = density in cm$^{-3}$, $A_{*}$ = density
scaling for the Wind-like profile so that $\rho = 5\times 10^{11}\,
A_{*}\,r^{-2}$~g$\;$cm$^{-1}$ ($r$ in cm; a standard reference for a
mass loss of 10$^{-5}\;$M$_{\odot}\;$yr$^{-1}$ at a wind speed of 1000
km$\;$s$^{-1}$), $E_{52}$ = isotropic-equivalent energy in units of
10$^{52}$~ergs, $t_d$ = observed time post-burst in days. The electron
energy partition $\epsilon_e$ is given as a fraction.

The post-$t_{jet}$ evolution with laterally expanding ejecta uses the
time-dependences of \citet{sph99} (eqs. 2-5), and is connected to the
pre-jet behaviour without smoothing (i.e., a sharp jet break).

Using the relativistic equations for the shock energy, $E=M\gamma^2$,
we calculate the time at which $\gamma=1$ as the nonrelativistic
transition, $t_{nr}$. This would be equivalent to using the
\citet{bm76} approximation for the energy, $E=M\gamma^2\beta^2$, and
defining the nonrelativistic transition condition as
$\gamma\beta=1$. We again employ a sharp transition to the
post-$t_{nr}$ behaviour.

We adjust this evolving synchrotron spectrum with self-consistent
corrections to the cooling rate, based upon the parameters already
enumerated. We calculate the effects of synchrotron photon upscatters
(inverse Compton scatters, IC) off the shocked electrons as in
\citet{se01}. When IC dominates the cooling we adjust the cooling
break $\nu_c$ (by $(1+Y)^{-2} \approx Y^{-2}$ when $Y>1$; $Y$ the IC
to synchrotron luminosity ratio $\approx$
(max[$(\gamma_c/\gamma_m)^{2-p}$,1]$\epsilon_e/\epsilon_B$)$^{1/2}$). The
upscattered photons can produce a high-frequency secondary source of
flux; for it we adopt the spectral breaks as in \citet{se01} and use
our synchrotron spectral shape (a good approximation except for an
ignored logarithmic correction between $\nu_m^{IC}$ and $\nu_c^{IC}$
and the slope for $\nu < \nu_a^{IC}$, where IC never dominates the
data).

We also treat radiative corrections self-consistently, from the model
parameters. Instantaneously, we treat the shock as adiabatic. The
energy is calculated for each time from the solution of $dE/E$
\citep{cps98}. It depends upon the ratio $\nu_c/\nu_m$ in slow
cooling ($\nu_c > \nu_m$) as the losses quench out and we use the
synchrotron-only rate of change in $\nu_c$, allowing a simple analytic
solution for $E(t)$. As this is when $E(t)$ changes are becoming
unimportant, the approximation has little effect. We scale the shock's
energy to the value at the change from fast- to slow-cooling regimes
($\nu_c=\nu_m$, fairly early).

This spectrum is modified on the way to the observer, and we must
account for extinction, interstellar scintillation (ISS) and the
host's flux. ISS in the Galaxy distorts the flux at low (radio)
frequencies; we estimate its fractional flux variations as outlined in
\citet{wal98} for point sources, using the map of scattering strengths
by \citet{tc93}, and scalings for extended sources as explained in
\citet{nar92}. The growing angular size quenches the scintillations; we 
assume initial parameters in the model to get angular size as a
function of time for the ISS model. These are not iterated, but used
as an additional uncertainty in the {\em model} flux, added in
quadrature to the data's uncertainties when estimating $\chi^2$.

For extinction, we first deredden for Galactic effects using
\citet{sfd98}'s method for estimating A$_{\lambda}$ in the optical. As 
most GRBs are at high Galactic latitudes, this term is small. Then, as
the extinction curves in high-$z$ galaxies are unknown, we use known
local extinction curves redshifted to the host restframe to fit for
host contributions. We use \citet{wd01}'s Large Magellanic Cloud (LMC)
in general, but for the GRB 980329 event we used the Small Magellanic
Cloud Bar's law due to evidence that the host's extinction law is
steep. We do not subtract host fluxes as the decay slope is sensitive
to small differences in the host value; host fluxes are added to the
model in the fit. In the optical and near-IR, when there is evidence
for a host component, we fit a value for the particular band(s)
involved.  As submillimeter hosts have been detected in other bursts
\citep{hlm+00, bkf01, bgcn1182+01, fbm+01, bck+03}, we allowed for
this possibility (scaled to 350 GHz as $\nu^3$), but none was required
for the four events under consideration and this component was not
included in the best fits.  Finally, we see in 2 datasets evidence for
underlying flux in the radio. 980703 has sufficient data that a
spectral index
\citep[-0.32,][]{bkf01} was determined and we fit the radio host flux
scaled to 1.43 GHz (where it is brightest). For 980329 we use a canonical
spectral index of $\nu^{-0.8}$, again scaled to 1.43 GHz.

Two other groups have performed full analyses similar to ours, as
described in \citet{cl00, cl01} and \citet{pk01, pk01b, pk02}. There
are some differences in each group's approach to the fireball
model. Chevalier \& Li develop an $r^{-2}$ CBM model, including
analytic solutions to the hydrodynamics, but they do a numerical
solution to get the full smooth spectral shape. They assume an extra
break in the electron energy distribution but do not, however, account
for energy losses, IC cooling or ISS effects. Panaitescu \& Kumar
numerically solve the dynamics (whereas we use analytic asymptotic
forms), numerically calculate the spectrum from equal arrival-time
surfaces (whereas we use equations adjusted to account for equal
arrival-time surfaces and the electron distribution behind the shock,
and a simple smoothing) and calculate IC scatters and energy losses
directly from the model's radiated spectrum (whereas we use the
theoretically expected levels from the parameters). They use the
simplest electron energy distribution when it provides a good solution
but also allow for two (rather than one) indices in the injected
spectrum if needed. They pre-account for host effects by subtracting
fluxes and fixing extinction levels, instead of including these in the
fits.

\section{The Sample of Four GRBs}\label{sec:data}

We selected four events with rich datasets (radio, optical-NIR and
X-ray at a minimum) for which the fireball model with simple
assumptions provides a good description of the lightcurves. These are
GRB~970508, detected by the {\em BeppoSAX} {\em GRBM} instrument, with
its X-ray afterglow found by the {\em BeppoSAX WFC} \citep{cfp+97c},
GRB~980329, its afterglow detected in the X-rays by \citet{iaa+98},
GRB~980703, detected by {\em BATSE} \citep{k+98} with its afterglow
identified by the {\em Rossi X-Ray Timing Explorer} \citep{lmm98} and
GRB~000926, located by the IPN \citep{hmg-gcn801, h-gcn802}, with its
afterglow first identified in the optical \citep{gcc+00, dfp+00}.

We compiled and described the afterglow data for the 980329, 980703
and 000926 events in previous papers. The summaries presented here
chiefly point the reader to the relevant papers. As we have not
compiled the 970508 data elsewhere, we provide full details here.

GRB~970508 was the second burst with a detected afterglow, and it
has a rich broadband dataset.  We
use the X-ray fluxes of \citet{paa+98}, converting to flux densities
and frequency assuming $\nu^{-1.1}$ over their quoted band. For the
optical/NIR, we use BVI data from \citet{skz+98} and \citet{zsb98b}, R
from the table compiled in \citet{gcm+98}, and K$_s$ from
\citet{cnm+98}. We converted magnitudes to flux densities with the 
information in \citet{tok00} for K$_s$ and \citet{b79} for BVRI. We
included the 12-$\mu$m mid-IR measurement by
\citet{hmd+99}. We used \citet{bbg+98} in the submillimeter (86 \&
232 GHz), and the 15 GHz data of \citet{tbfk97} and the 1.43, 4.86 and
8.46 GHz dataset of \citet{fwk00} in the radio. 

GRB 970508's afterglow was unusual
- a sudden brightening in (at least) the optical occurred at $\sim$ 1
day, followed by the usual powerlaw decays at $t \ge 2$ days, and the
dataset shows an intrinsic ``scatter'' even at later times in at least
the optical and radio. No simple model can account for the
rise. Various possible explanations for a sudden rise in afterglow
flux include a refreshed shock from slower material in the
relativistic flow catching up to the slowing forward shock
\citep{pmr98,sm00,kp00c}, a sudden jump in the encountered density 
\citep{wl00a,dl02,npg03}, 
or even the ``patchy shell'' model where there are hot and cold spots
in the shock and as the shell slows a hot spot may come into the
observable area \citep{npg03}. These models can provide a single rise that
will not affect the emission substantially later. To avoid such
complications we used only the data at $t \ge 2$ days, after the rise,
for the modelling dataset.

The 980329 event has data in the X-ray, optical, submillimeter and radio
regimes. Initial searches in the optical were unsuccessful, and
\citet{tfk+98} first identified the afterglow in the radio. Early epoch
optical data where the afterglow dominates over host emission is
therefore somewhat sparse. The excess $t < 3$~days emission at 8.46
GHz is thought to be due to a reverse shock \citep{sp99a} and
therefore we excluded it from our analysis. \citet{yfh+02} gives full
details concerning the dataset. No redshift has been determinable for
this event (the only such case in our sample) due to faint host
emission and a lack of prominent emission lines; we adopt z=2 here,
roughly the middle of the potential redshift range.

The 980703 afterglow data includes radio, optical/NIR (with good frequency
coverage: BVRIJHK) and X-ray. The utility of the opt-NIR data is
limited to a few days postburst by a bright host (which is also
visible in the late radio data). \citet{fyb+03} give the details.

The 000926 afterglow data includes radio data from 1.43 to 98 GHz,
ground-based and {\em HST} optical and NIR observations and {\em
Chandra} X-ray observations. The first 8.46 GHz point ($\approx$ 1.2
days postburst) was excluded from the analysis as abnormally high and
possibly associated with reverse shock emission. We use all available
BVRHK data and divide the X-ray into a soft and a hard band. This is
further described in \citet{hys+01}.

\section{Fits to the Basic Model}\label{sec:prevfits}

970508 shows a ``scatter'' in its lightcurves in all frequency regimes
and timescales. It is not clear how much of the scatter may be due to
real physical effects (such as clumpiness in the circumburst medium),
and how much could be attributed to cross-calibration
uncertainties. While we performed our fits as in \S\ref{sec:framework},
the best fit, passing through the data, has
$\chi^2$/DOF~$\approx$~2.5. (We do not attempt an analysis with
increased error bars to reduce $\chi^2$ to the DOF, as this would
require much larger uncertainties.) We have adopted this fit, a
nearly-isotropic solution ($t_{jet}=183$~days) with an ISM-like CBM of
somewhat low density (0.2 cm$^{-3}$) and $\epsilon_e$ and $\epsilon_B$
near equipartition, as our best solution.

\citet{yfh+02} fit 980329 as detailed in \S\ref{sec:framework}. Our best 
model is collimated, with a significant density (n $\sim$ 20
cm$^{-3}$) and host extinction. We found no need for a submillimeter
host component, and that a radio host improves the fit with marginal
significance. The best fit was not a unique solution; \citet{yfh+02}
noted an isotropic, extremely radiative ({\em all} the shock's energy
in radiating electrons) solution fit as well. We reject that solution
on the basis of its unphysical parameters.

\citet{fyb+03} gives results of 980703's fits as in \S\ref{sec:framework}.
The best fit is a collimated jet into a constant density CBM, its jet
break hidden in the optical by the host's dominance. It was not a
unique solution; a collimated jet in an $r^{-2}$ CBM fit the data
equally well. That fit required extreme parameters: 70\% of the
shock's energy is imparted to the electrons and the energy during the
afterglow is $\lsim$ 1/10 that of the prompt $\gamma$-ray emission,
requiring an extremely high $\gamma$-ray conversion efficiency. We
reject the $r^{-2}$ fit.

The model we employ (\S\ref{sec:framework}) for the 000926 fits has
changed since the work done in
\citet{hys+01}. Before, estimated host fluxes were subtracted instead
of fitted for. As there was a galaxy ``arc'' nearby contaminating the
ground-based data, the host components fitted for are merely the
persistent underlying flux common to both ground- and space-based
observations; the ``arc'' flux removal remains as before. Moreover,
the model now employs radiative corrections to the energy, which were
not yet incorporated in the previous analysis. We refit the dataset;
the best fit from the improved model is in Table
\ref{tab:bestfits}. It is still a high-IC ISM model with a substantial
density, but the $\chi^2$ has increased. The previous fit had a
nonnegligeable $\epsilon_e$ and no radiative corrections; with
radiative corrections correlated with $\epsilon_e$, these are
incompatible, $\epsilon_e$ drops and the fit worsens. In particular,
while the X-ray still requires flux above the synchrotron's level, the
parameter changes cause the IC flux estimate to no longer match the
data as well.

\subsection{Comparisons of the Four Bursts}\label{sec:fitresults}

\begin{figure}
	\epsscale{0.9}
  \plotone{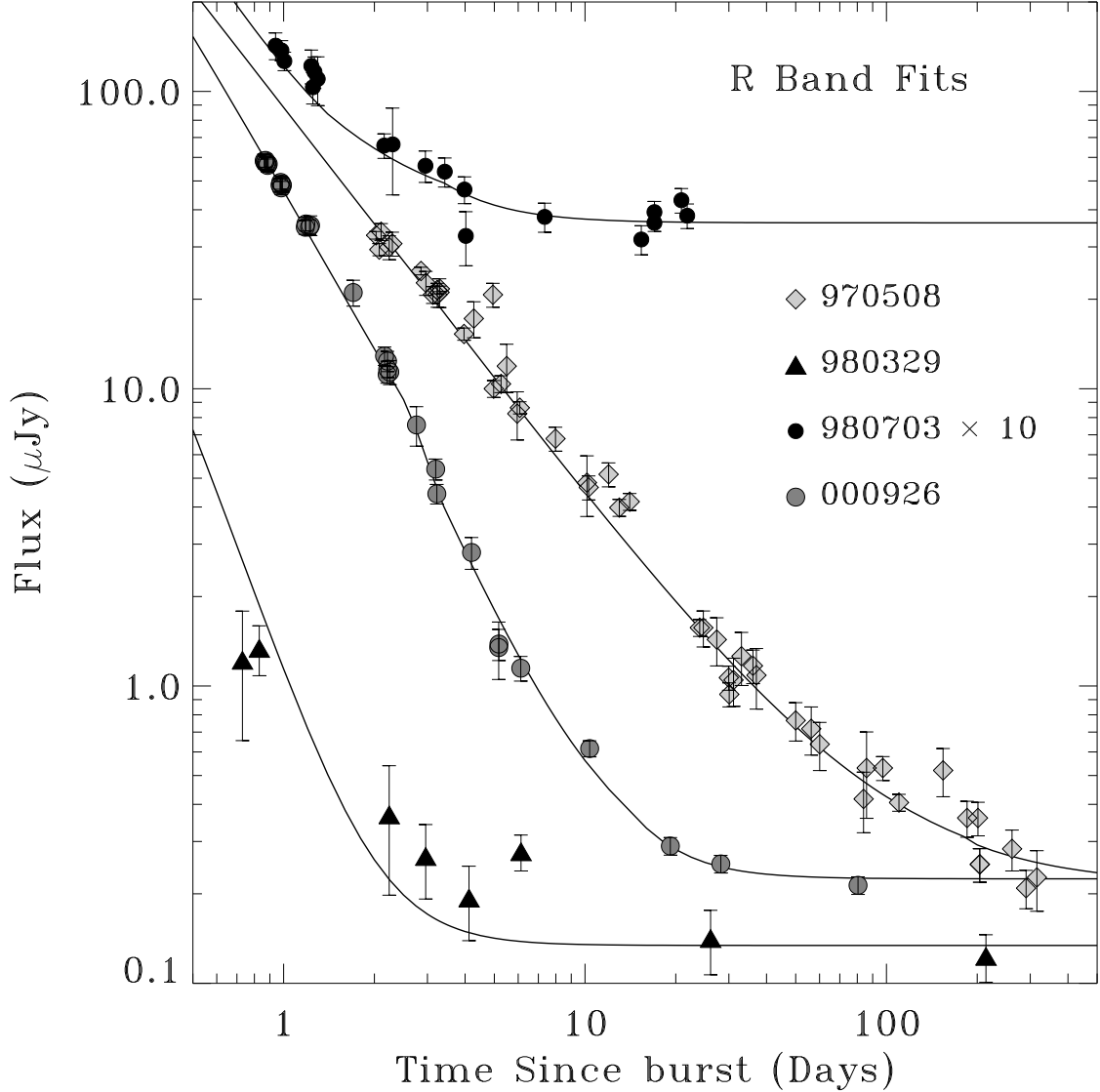}
\caption{
Subsets of the optical data from the best fits - the R band light
curves for the four events. For data selection, see
\S\ref{sec:data}. The fits shown are to the full broadband datasets,
with typically 100+ DOF, and are detailed in Table
\ref{tab:bestfits}. The GRB 980329 optical data is not very
constraining for the decay rate - a late optical detection (the early
points were found on reanalysis) did not allow deep followup over the
first week. The scatter in the GRB 970508 data cannot be explained in
any simple model, and otherwise the fits are quite good.
\label{fig_bestopt}}
\end{figure}

\begin{figure}
	\epsscale{0.9}
  \plotone{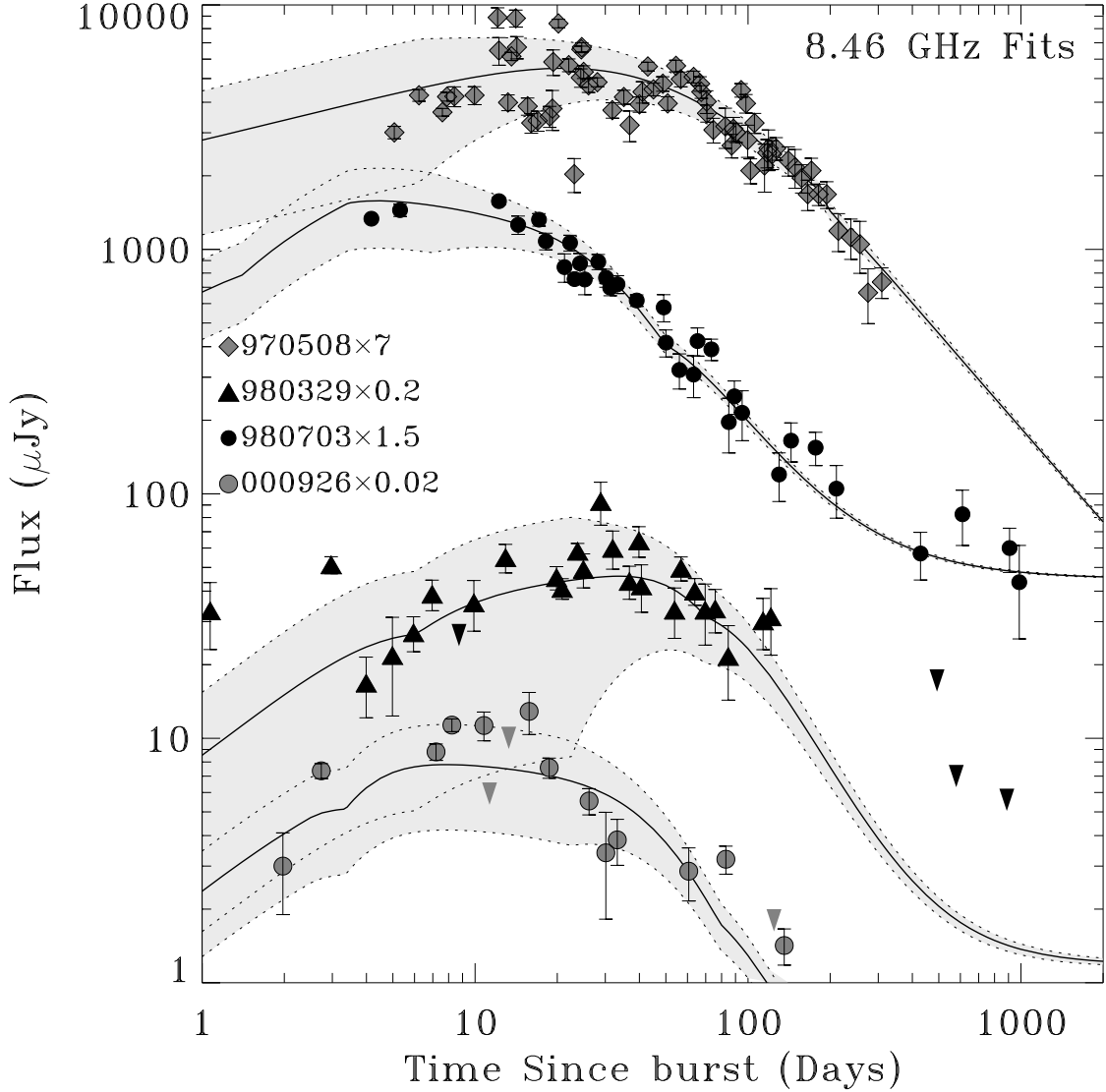}
\caption{
8.46 GHz light curves of the 4 events, with the best fits (see Table
\ref{tab:bestfits}). The light grey envelopes show the estimate of the
model uncertainty due to interstellar scintillation, and data that is
not statistically significant at the 2-$\sigma$ are presented as
2-$\sigma$ upper limits (downward triangles). As explained in
\S\ref{sec:data} the data for GRB 980329 prior to day four likely
contains an excess contribution from the reverse shock and was
therefore not included in the fit. The 980329 radio host component
only marginally improves the fit, due to a 1.43 GHz average excess,
which may indicate a weak radio host flux. The fits are overall quite
good, although ISS cannot fully account for the scatter in the 970508
dataset (there is, moreover, inexplicable scatter at other
frequencies).
\label{fig_best8GHz}}
\end{figure}

\begin{figure}
	\epsscale{1.}
  \plotone{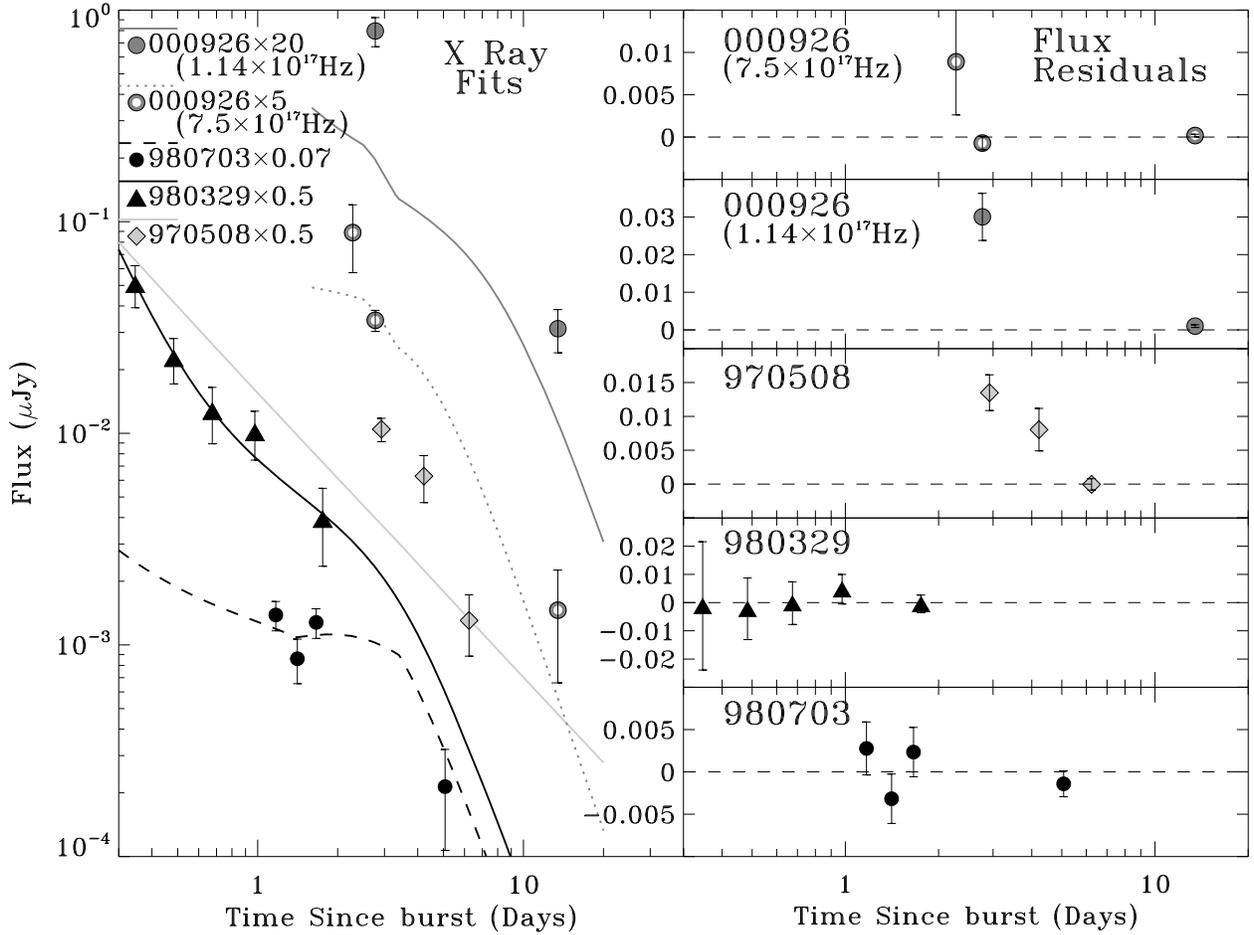}
\caption{
The X-ray data of the 4 events, with the best fits (see Table
\ref{tab:bestfits}). To parse out which model goes with each dataset,
the line style is indicated above the symbol labels, and
linearly-scaled residuals are on the right. The GRB 000926 data is
divided into a soft (solids) and a hard (open circles with dotted
line) band. The broadband fit (with relative flux levels and decay
rates) indicates an extra flux component in the X-ray, possibly
Inverse Compton upscatters, but our estimate of this component does
not completely fit the data. Better sampled X-ray lightcurves, such as
those expected from {\em Swift}, may clarify deficiencies in the X-ray
flux model. There is additionally a minor IC flux component in the GRB
980329 model, and IC dominates the early GRB 980703 X-ray model,
providing the flat initial flux and slow decay to match the data.
\label{fig_bestX}}
\end{figure}

\begin{figure}
	\epsscale{0.9}
  \plotone{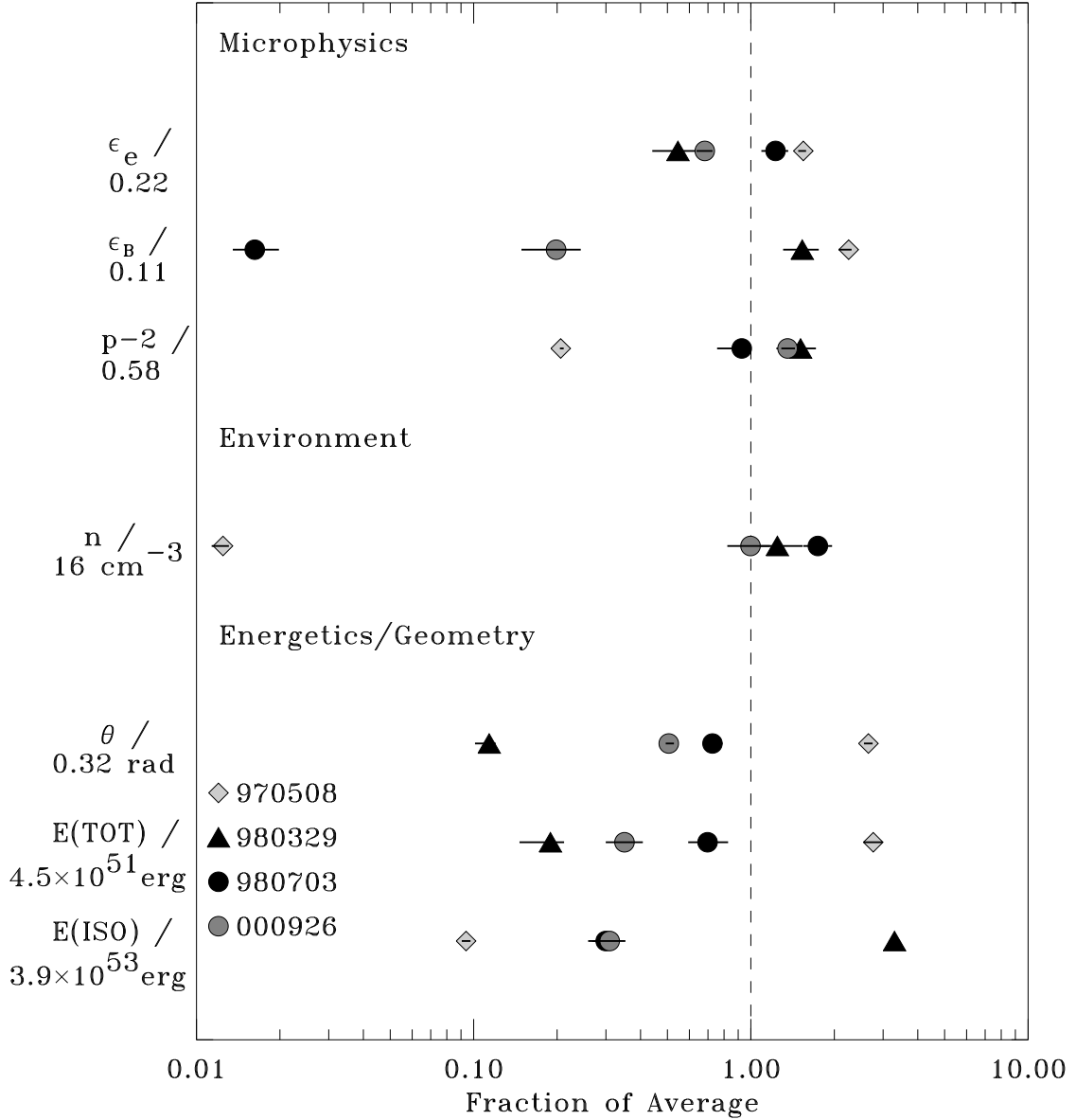}
\caption{
Parameters from the best fits with the simple model assumptions (see
Table \ref{tab:bestfits}). They are divided into 3 categories: energy
\& geometry (kinetic energy, collimation half-opening angle $\theta$),
environment (density) and microphysics (energy partitions:
$\epsilon_e$ for electrons, $\epsilon_B$ for magnetic fields, and
electron energy distribution index $p$). See \S\ref{sec:framework} for
further details concerning the fireball model's parameters. They are
presented relative to a nominal value as indicated; the error bars are
statistical only, 68.3\% intervals calculated via Monte Carlo
bootstraps. The model uncertainties are larger, as seen in the Tables'
range of parameters that produce reasonable fits under various
assumptions. The diversity in energy, geometry and environment is not
unexpected for some variation in progenitor properties. Shock physics,
however, is expected to depend only on shock strength. The variation
of these parameters by orders of magnitude suggests that some effect
is unaccounted for in the model.
\label{fig:bestpars}}
\end{figure}

Table \ref{tab:bestfits} presents the best fits using the basic
assumptions described in \S\ref{sec:basicmodel} (also see Figure
\ref{fig:bestpars}).  The table includes statistical 68.3\% confidence
intervals for the parameters, calculated from (non-Gaussian)
distribution histograms generated by over 1000 Monte Carlo bootstraps. The
results show a great deal of diversity in the values, similar to other
efforts \citep[e.g.][]{pk01}. The fits can be seen in Figures
\ref{fig_bestopt},
\ref{fig_best8GHz}, \ref{fig_bestX}.

Neither the variation in nor the values of the energy/geometry and
environmental parameters is surprising. Variation is seen in the
energies of the prompt GRB emission \citep{fks+01}. Moreover, if GRBs
are related to the deaths of massive stars
\citep[see the review by][and references therein]{m02} then we would
not expect such parameters to be identical due to variations in
progenitor mass, angular momentum and environment.

The densities are comparable to the Milky Way's ISM density at the low
end, and to the density of diffuse clouds for the three $\sim$ 20 cm$^{-3}$,
typical of other efforts with significant radio data (which constrains
the synchrotron self-absorption) \citep[e.g.,][]{pk01b}. The values
are not inconsistent with the core-collapse GRB progenitor hypothesis,
despite the expectation of a stellar birth environment, and molecular
cloud cores having $n \gg 10^{4}$ cm$^{-3}$. Diffuse clouds are also
associated with star-forming regions, and such densities are found in
the interclump medium that dominates molecular cloud volumes in the
Galaxy \citep[and its references]{c99}. Moreover, massive stars modify
their environment making direct interactions with the dense cloud cores
unlikely \citep{c99}.

The total kinetic energies inferred in the fits
(10$^{51}$-10$^{52}$~ergs) are roughly comparable, though at the high
end, with the range of total $\gamma$-ray energies expected from the
studies by \citep{fks+01}. On a case-by-case basis, we check if they
are reasonable by comparison with the $\gamma$-ray energies, employing
the isotropic-equivalent k-corrected 20-2000 keV fluences of
\citet{bfs01}. The ratio of the isotropic-equivalent kinetic (at 1 day 
post-burst) and $\gamma$ (over the GRB) energies varies from 0.5 to
5. However, the modelled energy corrections would give fireball
energies 3 to 10 times higher at the end of the burst; the implied
$\gamma$-ray efficiency would then be 3-10\% in three cases, and 40\%
for GRB~000926. The lower efficiency values are quite compatible with
theoretical expectations; 000926's is near the high end of the
predictions \citep[see][]{bel00,spm00,ks01}.

Thus the environmental and energy parameters are quite reasonable in
the present framework. What is somewhat unexpected is the lack of
universality in the microphysical parameters. Values of $\epsilon_e$
are fairly uniform, only varying by a factor of 2, but the values of
$\epsilon_B$ vary by a factor of 100. The index $p$ is expected to be
in the range of 2.2-2.3 \citep[see, e.g.,][]{kgg+00,agk+01} but its
values span from 2.1 to 2.9. While relativistic shocks are not
well-understood (neither fully modelled from first principles, nor
measured in the lab), the physics occurring at a shock boundary should
only be a function of the shock strength, or equivalently here for a
relativistic shock, its Lorentz factor. A spread by two decades in
$\epsilon_B$ would therefore be unexpected if everything is properly
accounted for in the model. We would expect the microphysical
parameters to be near-universal.

If the spread in the microphysical parameters does indeed exist then
highly relativistic shocks behave nonintuitively. (However, the
apparent spread could be due to the overall uncertainty in the model,
a parameter such as the viewing angle that is not fully accounted for
or nonuniqueness of the model fits.)  We have already noted that there
are at least two nonunique fits: GRB~980329 with a highly radiative
fit, and GRB~980703 with an $r^{-2}$ density profile fit disfavoured
due to extreme parameters. This leaves open the question as to whether
there may be equally good fits with reasonable parameters (perhaps
under other model assumptions). It is therefore important to check if
the model, with the very simple assumptions we have employed, is
constrained only to these fits by the data. The assumptions could be
too simple, and the following section explores fits with other model
assumptions in order to constrain some interesting aspects of the
model uncertainty.

\section{Constraints on Deviations From the Basic Model}\label{sec:dev}

Above, we found good fits even with simple assumptions. But
without quantifying the uncertainty in model assumptions, it is not
clear if this says anything about the simplicity of the underlying
physical processes or the environment. We do not know if fits are
possible with a variety of underlying assumptions, and whether that
would substantially change the parameters inferred. We now investigate
the range of allowed modifications on a limited set of fireball model
assumptions, and the effect of modified assumptions on the parameters
in good fits.

First, the magnetic field amplification (assumed to be from
instabilities) is not well understood - we do not know from first
principles whether the fields pile up, decay away or reach a steady
state. There is also a spread in the magnetic energy fraction
$\epsilon_B$ (by a factor of more than 100). So the behaviour of the
magnetic fields is clearly one of the important questions surrounding
the microphysics of relativistic shocks. We want to learn whether it
is possible for the field energy to evolve with the strength of the
shock in the afterglow phase, and if the data truly constrains the
interesting diversity seen in the best fit values. To get some
constraint on this \S\ref{sec:epbofx} explores whether an evolving
$\epsilon_B$ could fit the data.

Second, we know there is at least one case where more than one density
profile can produce a good fit (\S\ref{sec:prevfits}). There is also a
growing body of evidence that GRBs are associated with star-forming
regions and may be the result of the collapse of massive stars
\citep[see][and references therein]{m02}. If correct, the shock should be
propagating into a medium enriched by the stellar wind of the
progenitor, and the density profile should not be constant. However,
the best fits are for a constant-density medium. We want a constraint
on how sensitive the fits are to density profile, and so we investigate a
wider variety of density profiles ($\rho \propto r^{S}$) than the two
previously considered, in
\S\ref{sec:n_ofr}.

\subsection{Magnetic Energy Fraction as Function of $\gamma_{shock}$\label{sec:epbofx}}

The strength of the magnetic fields implied in the fireball model are
far stronger than the levels expected from the strength of the
Galactic fields. The relativistic shock is expected to amplify the
nearby fields \citep[e.g.][]{ml99}. This would be a nonlinear process
acting on small instabilities; the resulting fields would depend on
the amplification mechanism and not on the initial values. This aspect
of relativistic shock physics is not fully understood
\citep[see the review by][and references therein]{kd99}. GRB afterglow 
emission can be used to investigate this process; \citet{rr03} show
that spectra will differ if the magnetic field is confined to a small
length behind the shock. The importance of magnetic fields in GRBs is
further demonstrated by \citet{cb03}'s recent detection of
linear polarization at the theoretical maximum in the prompt emission
of GRB 021206. This would require a uniform magnetic field across the
$\gamma$-ray emission region beamed to the observer, potentially
implying that magnetic fields dominate the dynamics \citep{lpb03}.

The assumption that the magnetic energy fraction $\epsilon_B$ imparted
by the shock amplification is constant is the simplest assumption;
shock physics could depend on the Mach number (equivalently the
Lorentz factor if relativistic). Therefore it is possible that the
shock would have different efficiencies with time, and the fitted
number is simply some averaging over where its modelled effect on the
data is strongest. It would also be possible to get effectively an
increase or decrease in $\epsilon_B$ if the pileup of magnetic fields
behind the shock in the place where the electrons are accelerated and
radiate does not reach a steady state - if the fields remain and grow
it would increase and if they diffuse away faster than they are
replenished it would decrease. (It is also therefore possible that the
pileup could be position-dependent, even if there is a universal
behaviour for a collimated flow, and some disparities could be due to
the unaccounted-for differences in viewing angle.)

It is important to constrain whether the data {\em allows} a variable
$\epsilon_B$ to see if such explanations are viable, and its spread
could be due to model uncertainty. If the $\epsilon_B$ disparities
could be resolved with a universal nonconstant $\epsilon_B$ behaviour
that would be of significant interest. Moreover, if $\epsilon_B$'s
behaviour is nonconstant, the values fitted with the simple constant
assumption could be different from the real magnetic
fraction; the level of this effect should be checked. And if only a
growing $\epsilon_B$ (pileup of field energy as the shock strength drops) or
a falling $\epsilon_B$ (drop in the field energy as the shock strength
drops) are allowed by the data, this would be a clue to the unknown
details of relativistic shock's field amplification. To see what the
data permit, we studied a simple parametrization of $\epsilon_B$ with
the shock's $\gamma$.

We took the equations for the model as presented in
\S\ref{sec:framework} and allowed the fixed magnetic energy fraction to
vary smoothly. A value for the parameter is calculated at each time
and inserted into the equations governing the spectrum at that
time. Under assumptions where $\epsilon_B$ grows we do not permit it
to get infinitely large, capping it at 100\%. Any change in the
magnetic energy fraction is expected to come from the evolution of the
shock, and thus the simplest physically motivated form is to tie it to
the shock strength, as expressed by its Lorentz factor $\gamma$. For
simplicity, we considered $\epsilon_B$~$\propto$~$\gamma^x$, where the
basic model (\S\ref{sec:basicmodel}) has $x=0$.

This affects the results in two ways, directly and indirectly. There
is an explicit dependence upon $\epsilon_B$ in the spectral breaks
\citep{spn98} whose evolution changes. Indirectly, as the cooling
frequency affects the energy losses as described in
\S\ref{sec:emission}, we also recalculate $E(t)$, which affects the
dynamics.\footnote{Note that for $x<-4/3$ synchrotron-only theory
predicts a decreasing $\nu_c/\nu_m$ requiring us to change the
function $E(t)$. The transition to fast cooling, where all injected
electrons can radiate a significant part of their energy, would only
permit energy losses to quench out once $\epsilon_B$ stops changing -
as late as the nonrelativistic transition. In practice this is not the
case in models of interest with $\epsilon_B$ rising over the data
range; $\epsilon_e$ is large enough that $\epsilon_e > \epsilon_B$. IC
cooling dominates so $\nu_c \approx \nu^{synch}_c\times\epsilon_B/\epsilon_e$
\citep{se01}. With moderate energy losses at even $x=-3$ the ratio rises 
at a low rate. The radiative corrections fall between no quenching of
the energy losses and the quenching rate of $x=0$. These are not very
different; typically the energy difference between these out to $\sim
200$ days is $\sim 2\times$, a small effect on the final data with low
S/N. We use the synchrotron $\nu_c/\nu_m$ ratio to calculate quenching
above $x=-4/3$ and check both no quenching of $E(t)$ and full $x=0$
quenching for $x<-4/3$. These give substantively the same results.}

The resulting equations follow ($\nu_a$ is for the order
$\nu_a~<~\nu_m~<~\nu_c$. When $\nu_c~<~\nu_m$, multiply by
$(\nu_m/\nu_c)^{1/2}$).

\begin{equation}
\begin{array}{lcccc}
 & & t<t_{jet} & t_{jet}<t<t_{nr} & t_{nr}<t \\
\hline
F_{max} \propto & E n^{0.5} \epsilon_e^0 \epsilon_{B}^{0.5} & t^{0-0.19x} & t^{-1-0.25x} & t^{0.6} \\
\nu_a \propto & E^{0.2} n^{0.6} \epsilon_e^{-1} \epsilon_{B}^{0.2} & t^{0-0.08x} & t^{-0.2-0.1x}& t^{1.2} \\
\nu_m \propto& E^{0.5} n^{0} \epsilon_e^{2} \epsilon_{B}^{0.5} & t^{-1.5-0.19x} & t^{-2-0.25x}& t^{-3} \\
\nu_c \propto& E^{-0.5} n^{-1} \epsilon_e^{0} \epsilon_{B}^{-1.5} & t^{-0.5+0.56x} & t^{0+0.75x}& t^{-0.2} \\
\end{array}
\end{equation}

We examined model assumptions of several indices $x$, $x=$~2, 1, 0,
-1, -2, -3. We did not free-fit the index, considering instead integer
values. Changes are not highly sensitive to the index; for a typical
time span of a factor of 100, $\gamma$ changes by $\approx$1/6-1/10, and
the change in $\epsilon_B$ for $x=1$ alters the magnetic field by only
a factor $\approx$2.4. Table \ref{tab:epbx_fnu} gives the expected
behaviour of the synchrotron flux for $x \ne 0$; these changes are not
large for $\delta x \ll 1$.

We thoroughly searched for the best fit models at the values of $x$
considered. Our methods included stepping $x$ in very small steps
(between 0.05 and 0.2, depending upon the ease with which the fitting
would adjust for a dataset), allowing small, smooth changes. We tried
larger steps $\delta x$ ($\sim 0.1$ where $0.05$ would work smoothly,
$\sim 0.25$ or $1$ where $0.2$ would work for small steps), forcing
the gradient search algorithm to look farther afield for a good
fit. We also fit from grids of selected parameter starting points at a
particular (integer) $x$. These grids included values comparable to
those at the best fit for the $x$ next-nearest to 0, the basic
model's values and typically went up \& down by a factor $\sim 10$.

\begin{figure}
	\epsscale{0.85}
  \plotone{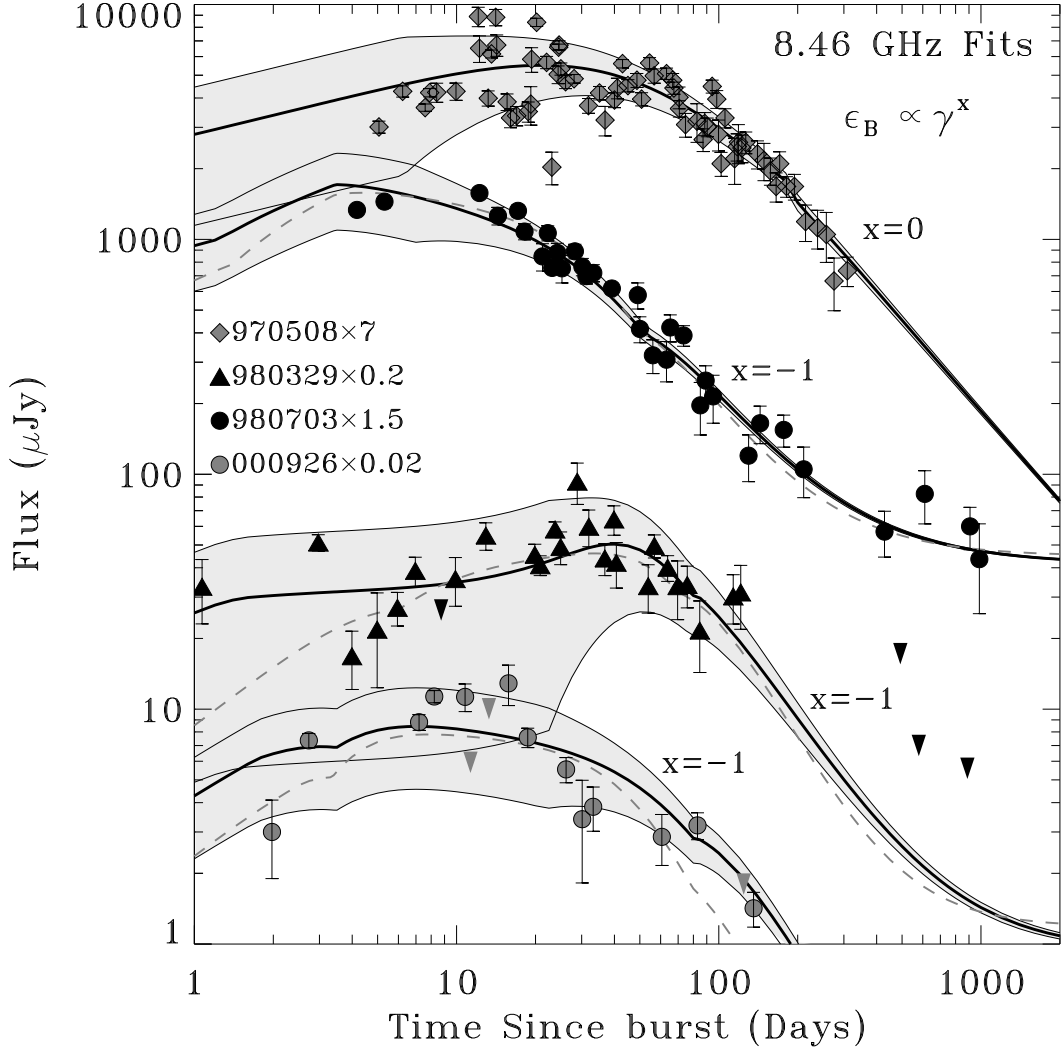}
\caption{
8.46 GHz light curves of the four events, with the best fits (solid
lines) for an assumed magnetic energy relation $\epsilon_B \propto
\gamma^{x}$ (\S\ref{sec:epbofx}). The light grey envelopes are the 
estimated scintillation uncertainties (the 970508 scatter is not fully
accounted for, but it has scatter excess at all frequencies); data
that isn't 3$\sigma$ significant is shown as 2$\sigma$ upper limits
(downward triangles). 980329 has a radio host component which
marginally improves the fit due to a 1.43 GHz average flux excess. In
one case (970508) a constant $\epsilon_B$ produces the best fit (but
only by 1\% in the total broadband $\chi^2$); in the others
$\epsilon_B \propto \gamma^{-1}$ gives the best fit. For these three, the
best constant-$\epsilon_B$ fit is shown as a grey dashed line for
comparison. The model decay for constant-$\epsilon_B$ is generally
slightly steeper than the data in the late radio; this is especially
obvious in the last few points of the 000926 light curve. A magnetic
energy increase at late times flattens this decay, improving the fit.
\label{fig:bestx.8GHz.comp}}
\end{figure}

\begin{figure}
	\epsscale{0.85}
  \plotone{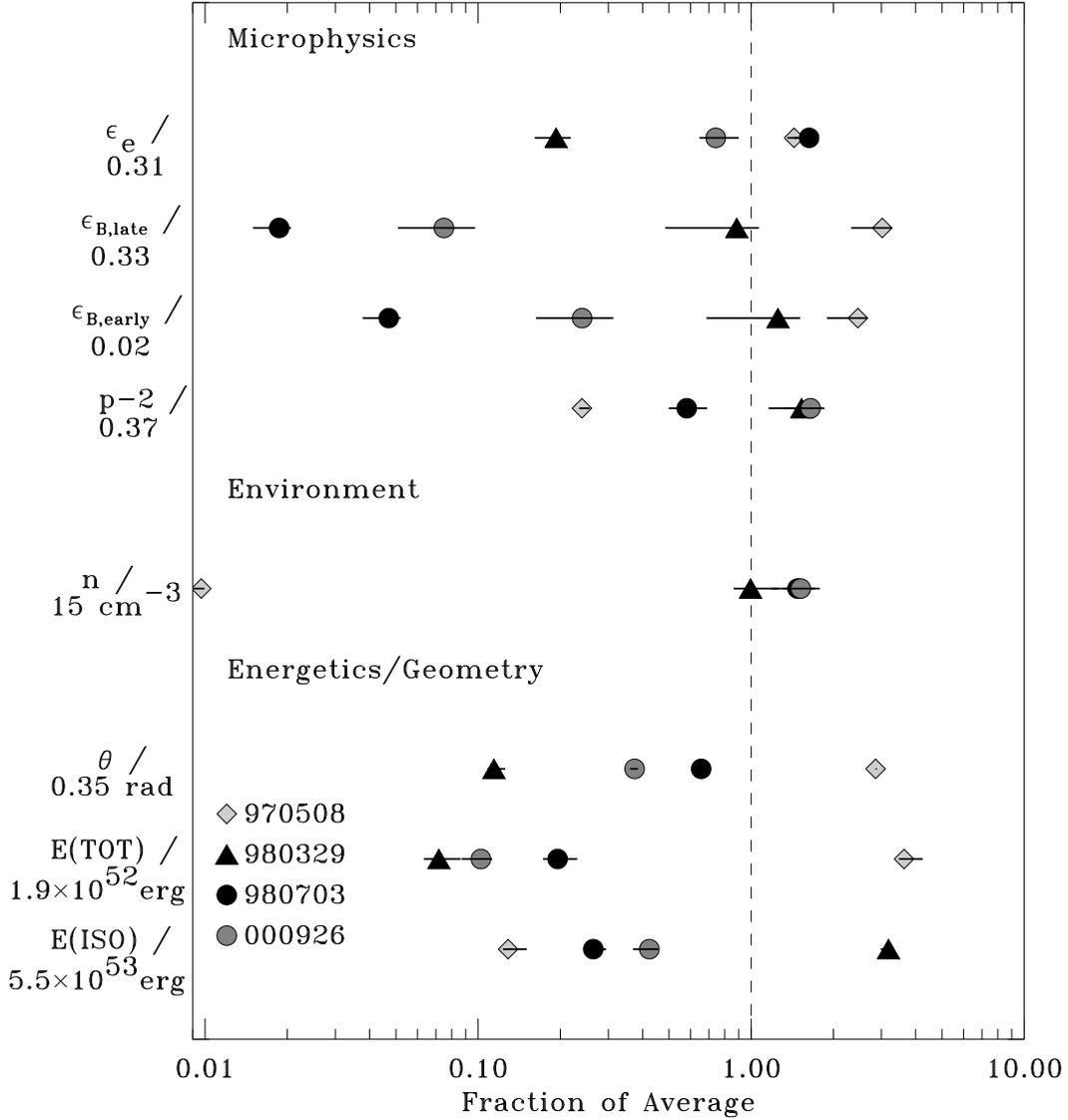}
\caption{
Parameters from the best fits assuming $\epsilon_B \propto
\gamma^{-1}$ (see \S\ref{sec:epbofx}, Table \ref{tab:bestxn1}).  They
are divided into 3 categories: energy \& geometry (kinetic energy,
collimation half-opening angle $\theta$), environment (density) and
microphysics (energy partitions: $\epsilon_e$ for electrons,
$\epsilon_B$ for magnetic fields, and electron energy distribution
index $p$). As it varies, $\epsilon_B$ is presented both with its
early (at fast- to slow-cooling transition) and final values.  See
\S\ref{sec:framework} for further details concerning the fireball
model's parameters. They are presented relative to a nominal value as
indicated; the error bars are statistical only, 68.3\% intervals
calculated via Monte Carlo bootstraps. The model uncertainties are
larger, as seen in the Tables' range of parameters that produce
reasonable fits under various assumptions. This model assumption
produces fits as good as (970508) or better than (980329, 980703,
000926) those assuming $\epsilon_B$ is constant. However, it does not
fit with a universality in the microphysics. There is clearly great
flexibility in the model assumptions allowed by the data, and
considerable {\em model} uncertainty in the derived parameters.
\label{fig:xn1pars}}
\end{figure}

Best fits for various $x$ are in Table \ref{tab:epbx_bestfits}. We
find that $x<0$ fits as well or better than the basic model out to at
least $x=-1$; the best $\epsilon_B \propto \gamma^{-1}$ fits including
uncertainties are shown in Table \ref{tab:bestxn1} and Figure
\ref{fig:xn1pars}. The improvement is seen in 3 of the 4 cases, while
the fit with the $x=-1$ model assumption to 970508 is only 1\% worse
in the total broadband $\chi^2$ than the basic, constant-$\epsilon_B$
fit. The effect of the $x=-1$ assumption on the fit to optical and
X-ray data is minor. It takes a long time baseline, as in the radio,
to see the spectral evolution differences; moreover the optical and
X-ray are above the cooling frequency and thus have a lesser flux
dependence upon $\epsilon_B$ (Table \ref{tab:epbx_fnu}). Assuming
$\epsilon_B \propto \gamma^{-1}$ improves the radio fits - the
magnetic energy grows as the shock slows, which allows the peak to
rise, flattening the late time decay, as shown in Figure
\ref{fig:bestx.8GHz.comp}.  We note there appears to be a general
trend that the radio decays are a bit shallower than the optical /
X-ray, so an increasing magnetic energy improves the fit. There are
other possible causes; any effect that increases the peak flux, such
as an increasing energy or density, will flatten the radio decay.

For a sufficiently steep $\epsilon_B \propto \gamma^{x}$; $x>0$
the resulting model behaviour can no longer fit the datasets despite
any compensating changes in other parameters. As the peak flux
$\propto \epsilon_B^{1/2}$, one of the chief spectral behaviour
changes is that the peak flux drops, producing steeper decays (see
Table \ref{tab:epbx_fnu}). This leads to a poor
radio fit by $x \approx +3$, often with the radio peak flux too low in
order to give an appropriate earlier peak flux relative to the optical
data.
The effect is stronger post-jet, and as shown in Table
\ref{tab:epbx_bestfits} the data for 000926 (with a clear jet break in
the optical $\approx$~2 days) does not even produce a good fit
assuming $x=+1$. The extra steepening in the radio requires a later
jet break than is compatible with the optical; attempts were made to
compensate in fits with $t_{jet} \lsim 3$ days, but the radio could
not be forced to fit. There can also be some fit difficulties at
higher frequencies; these may be due to a drop in IC flux components
for higher early $\epsilon_B$, or a change in the spectral slope from
its link to the decay rates.
These effects also combine; a decreased IC flux in the X-ray requires
an increased synchrotron flux. In some models it is produced by
overproducing the model's optical flux and suppressing it with host
extinction. The combination does not produce appropriate spectral
indices both in the optical and from the optical to the X-ray.

While the $x=-1$ assumption produced improved fits, for a negative
enough $x$, $\epsilon_B \propto \gamma^{x}$ can no longer match the
data, at $x \approx -3$. For $x \ll 0$, $\epsilon_B$ will become
unphysically high, capping at 100\%, even with a reasonable early
value. This occurs at $x=-4$ for 980329, where no good fit can be
found with $\epsilon_B < 100\%$ in the data range; early radio flux
would decline rather than rising, in contradiction to the data (see
Table \ref{tab:epbx_fnu}). However, the others cannot fit before
reaching that extreme. By $x=-3$, 980703's radio peak cannot be
matched.
The peak frequency drops so slowly that it does not pass until quite
late; a decline is produced post-jet (so too early) by pushing
$\epsilon_e \rightarrow 1$ to maximize the energy losses. This
unphysical result is also required to counteract the shallow decay at
optical and X-ray frequencies.
970508 fits poorly in a similar manner, but
000926's poor fit is again due to the definite optical break. With IC
cooling dominant, the pre-jet flux is too shallow, so the jet break is
pushed back earlier than the data, with a break due to the steepening
associated with the change to synchrotron cooling dominance which does
not match the data well.

In summary, the data does not constrain the fireball model to have a
constant magnetic energy fraction $\epsilon_B$. Good fits are possible
with both increasing and decreasing $\epsilon_B$, generally as
strongly as $\epsilon_B \propto \gamma^{+1}$ through $\gamma^{-2}$. With
$\gamma$ changing by a factor $\sim$ 10 over the data range, this
allows an extra change in magnetic field strength by a factor of
$\sim$3-10. Moreover the increasing magnetic energy with $\epsilon_B
\propto \gamma^{-1}$ tends to improve the model fit to the radio data. 
As $\vec{B} \propto \gamma \epsilon_B^{1/2}$, a more constant level 
$\vec{B} \propto \gamma^{1/2}$ fits slightly better than the basic model.

\subsection{An Investigation of Density Profiles $n \propto r^{S}$}\label{sec:n_ofr}

The CBM density distribution is an important clue to the nature of the
progenitor, with an $r^{-2}$ profile expected as the signature of a
massive star \citep[e.g.][]{dl98, cl00}. A number of profiles have
already been considered. This includes ``naked GRBs'' where a constant
density drops to zero after a certain radius
\citep{kp00b} and \citet{rdm+01}'s evolution of the afterglow through
the ejecta left by a Wolf-Rayet star showing that $r^{-2}$ is a very
crude approximation to the environment about a massive
star. Nevertheless, most model fits do as well or better with the ISM
approximation to the density profile than the Wind one
\citep[e.g.,][]{pk02}, although there is one case where a Wind CBM fits and 
an ISM cannot \citep{pbr02}.

We investigate a wide range of density profiles, parametrized
as $n \propto r^{S}$. We recalculated the equations for the model
presented in \S\ref{sec:basicmodel} allowing for a general powerlaw
density profile index $S$. This includes the changes to the radiative
loss estimates, as in \citet{cps98}. There are no general calculated
adjustments in the equation normalizations accounting for the
postshock electron distribution with a generic density profile, so we
used the order of magnitude estimates (including for $\gamma$, which
gives $t_{jet}$), and the estimate that from the jet expansion
$t_{NR}/t_{jet} \approx \theta^2$.

We expect cases of $S<0$ to help constrain how sensitive the model is
to the details of a mass-loss wind from the supposed progenitor
star. Cases of $S<-2$ approach that of an evacuated cavity. The
equations for the fireball energy from \citet{bm76} break down at
$S=-4$ \citep[see][for $S<-4$]{bs00}.

Conversely, $S>0$ models a fireball plowing into a medium that
gradually increases in density. This could mimic a denser region
surrounding the burst (though not a sharply bounded overdensity). For
$S \gg 1$, it offers insight into the behaviour when hitting the edge
of a very dense, but not sharp, shell surrounding the burst. It does
not directly model the extinction column from the material, so the
density would have to cut off before the material ahead would absorb
all the light emitted from the shock.

Following is the the expected behaviour of the spectral breaks with
general $S$. $\nu_a$ is for the order $\nu_a~<~\nu_m~<~\nu_c$. When
$\nu_a~<~\nu_c~<~\nu_m$, it is multiplied by $(\nu_m/\nu_c)^{1/2}$;
other orderings have other factors.

\begin{equation}
\begin{array}{lcccc}
 & & t<t_{jet} & t_{jet}<t<t_{nr} & t_{nr}<t \\
\hline
F_{max} \propto & E^{(8+3s)/(8+2s)} n_i^{2/(4+s)} \epsilon_e^0 \epsilon_{B}^{0.5} & t^{s/(8+2s)} & t^{-1} & t^{(2s+3)/(s+5)} \\
\nu_a \propto & E^{0.8(s+1)/(s+4)}  n_i^{2.4/(s+4)} \epsilon_e^{-1} \epsilon_{B}^{0.2} & t^{0.6s/(s+4)} & t^{-0.2} & t^{(16s+30)/(5s+25)} \\
\nu_m \propto & E^{0.5} n_i^{0} \epsilon_e^{2} \epsilon_{B}^{0.5} & t^{-1.5} & t^{-2} & t^{-(4s+15)/(s+5)} \\
\nu_c \propto & E^{-(3s+4)/(8+2s)} n_i^{-4/(s+4)} \epsilon_e^{0} \epsilon_{B}^{-1.5} & t^{-(3s+4)/(8+2s)} & t^{0} & t^{-(2s+1)/(s+5)}\\
\end{array}
\end{equation}

An increasing density ($S>0$) does not change the behaviour as greatly
as a rarefaction ($S<0$). With the converse, rapid changes are
expected for decreasing $S<0$, not increasing $S>0$. The reasons are
illustrated by the range in densities probed and the rate at which the
shock $\gamma$ slows depend upon a density profile.

Taking the observer time $t$ affected by relativistic beaming, and the
relativistic energy, $E=M\gamma^2$ (c=1), with $M$ from the density
profile $n \propto r^S$ the results are \[E \propto r^{S+3}\gamma^2,
\; t \propto r\gamma^{-2}\] Solving for $r$ gives 
\[n \propto r^S \propto (Et)^{S/(S+4)}\]
As evident from the above equation, for $S>$~a few the rate at which
density $n$ increases with {\em observer time} is weakly dependent
upon $S$, despite its strong dependence on radius. As well, in the
adiabatic approximation, with $E$ constant
\[\gamma \propto t^{-(S+3)/(2S+8)}\]
For $S \geq 0$, the rate of slowing only varies from from $\gamma
\propto t^{-3/8}$ to $\gamma \propto t^{-1/2}$; it depends quite
weakly on the density profile for $S$ larger than a few.

Moreover, in our model, the post-jet shock evolution is not sensitive
to density (since it is stopped and expands laterally) until the
nonrelativistic transition. Changes in the parameters to maintain the
same jet break for different $S$ may make the nonrelativistic
transition come at a somewhat different time, but the effect is not
dramatic. The question becomes whether the model with a given $S$ can
reproduce the behaviour pre-jet. To examine this, we choose the two
datasets with sufficient data for well constrained jet break
times. GRB 970508, which was well fit with near-isotropy, and GRB
000926, with a sharp jet break seen in the optical, are most useful to
constrain possible model fits for various $S$. The others had more
limited optical data, where jet breaks are generally most obvious, as
it was either sparsely sampled or partly masked by a bright host. (In
\S\ref{sec:prevfits} we note that the GRB 980703 dataset could be fit
equally well by ISM and Wind-like density profiles; clearly at least
some very good datasets do not constrain $n(r)$.)

We have seen cases where $n \propto r^{-2}$ densities can fit the
data, but it is difficult to even marginally fit models with as steep
a gradient as $r^{-2.5}$ (see the best fits in Table
\ref{tab:nofr_fits}). For 000926,
we could not find a satisfactory fit with $r^{-2.5}$, as no good fit
was possible with a jet break.
The peak flux declines rapidly prejet; if it scales to fit the early
optical it is low by the jet break, then too low in the radio.  The
radio model's decay begins too early due to $\nu_a$'s rapid decline as
the shock moves into less dense material.
GRB 970508 is also not well fit by $n \propto
r^{-2.5}$. Its $r^{-2.5}$ model with the best $\chi^2$ in Table
\ref{tab:nofr_fits} has an unphysical $\epsilon_B=100\%$, giving high 
initial spectral breaks $\nu_a$ and $\nu_m$ as these drop quickly. The
peak is high early (as it drops quickly); the spectrum from the peak
to the optical frequencies is steeper than in the ISM case. That
steepness out to the X-ray frequencies causes that model to
underpredict the X-ray flux.
The best 970508 model for $r^{-2.5}$ under the constraint that
$\epsilon_B < 100\%$ does not fit the radio data; its predicted flux
rises and falls too sharply as $\nu_a$ and $\nu_m$ fall rapidly.

As mentioned above, for an increasing density gradient the shock slows
rapidly compared to the constant density case, and the resulting
changes in $n$ seen by the shock are gradual.  As a result, 970508 and
000926 can both be fit with $S \approx 10$ (Table
\ref{tab:nofr_fits}).  For 000926, the increasing density leads to a
more rapid nonrelativistic transition, improving the fit somewhat
relative to the constant density case.  Further changes will go too
far; the fit becomes marginal around $S=12$ where the best fit models
overestimate the early radio flux.  970508 has no visible jet break,
and is not sensitive to $S$ even for values exceeding $S \gsim 10$.
Pushing to extreme $S$ to discover how far the assumptions can fit the
970508 data (eventually the gradient would make the nonrelativistic
transition come too early) is of limited value; GRB 970508's data is
not very sensitive to an increasing CBM density gradient.

In summary, while the data may sometimes accomodate an $r^{-2}$ CBM,
it does not fit an extreme blown-out density $r^{-2.5}$. But a shock
plowing into a denser region cannot be easily excluded by the
data. This is not the same as a sudden jump in density, but a gradual,
continuous increase $n \propto r^{S}$, for $S \gg 1$, which is not very
realistic but may roughly mimic a dense but not sharp shell of
material (perhaps ejecta from the progenitor). We conclude that the
fireball model data fits are not very sensitive to {\em increasing}
density gradients.

\section{Improving Observational Constraints}\label{sec:betterobs}

\begin{figure}
	\epsscale{0.9} \plotone{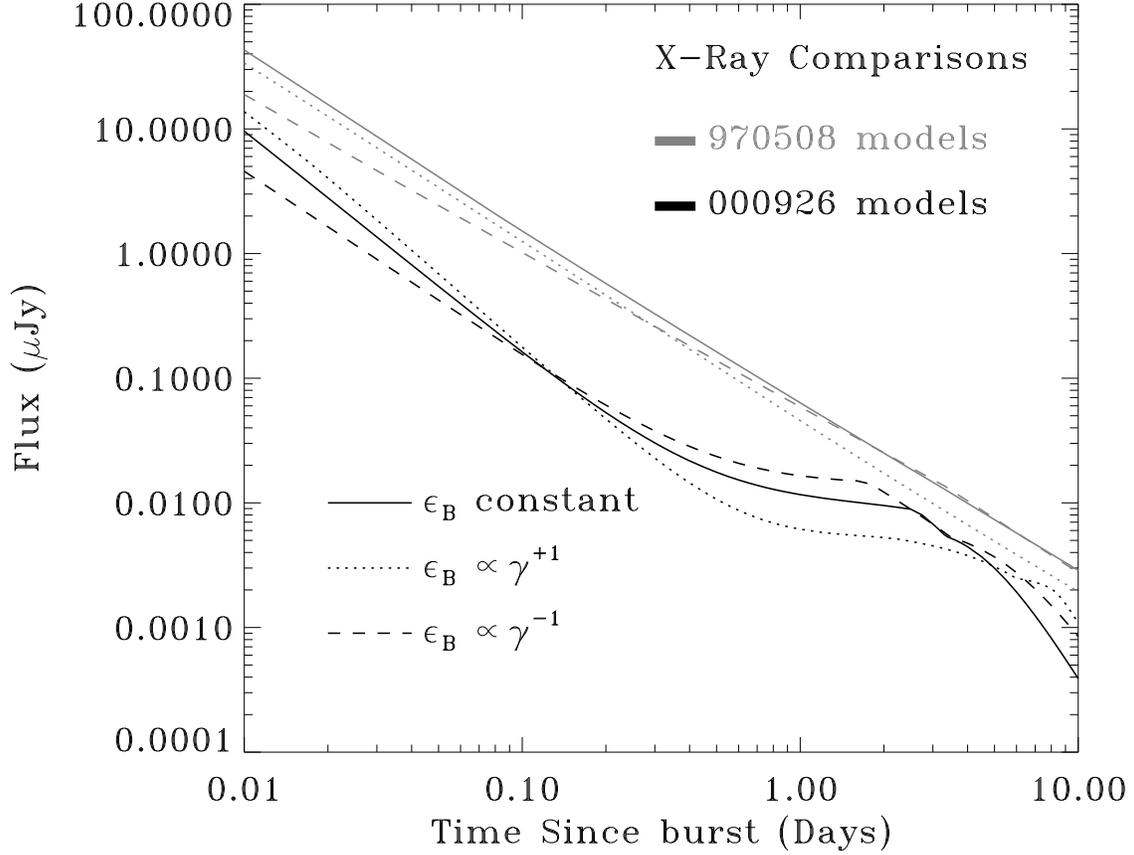}
\caption{
Comparison of model X-ray lightcurves for several equally acceptable
model fits. The lightcurves are for a frequency of $6 \times
10^{17}$Hz (nominal for {\em Swift}), and show the good fits for GRBs
970508 and 000926 with the basic model as well as with the assumption
that the magnetic energy fraction $\propto \gamma^{\pm 1}$. While the
models are close around the times of the X-ray observations ($\sim$
days), they diverge at early times, with a spread of several $\mu$Jy
at 0.01-0.03 days. Moreover, in the case of 000926, there is
significant IC upscattered flux, which gives different peak passage
times under the differing model assumptions. Early, sensitive, densely
sampled lightcurves such as those expected from {\em Swift} would
determine if we are modelling the X-ray flux, including the
upscattered IC photons, correctly.
\label{xraycomps}}
\end{figure}

\begin{figure}
	\epsscale{0.9} \plotone{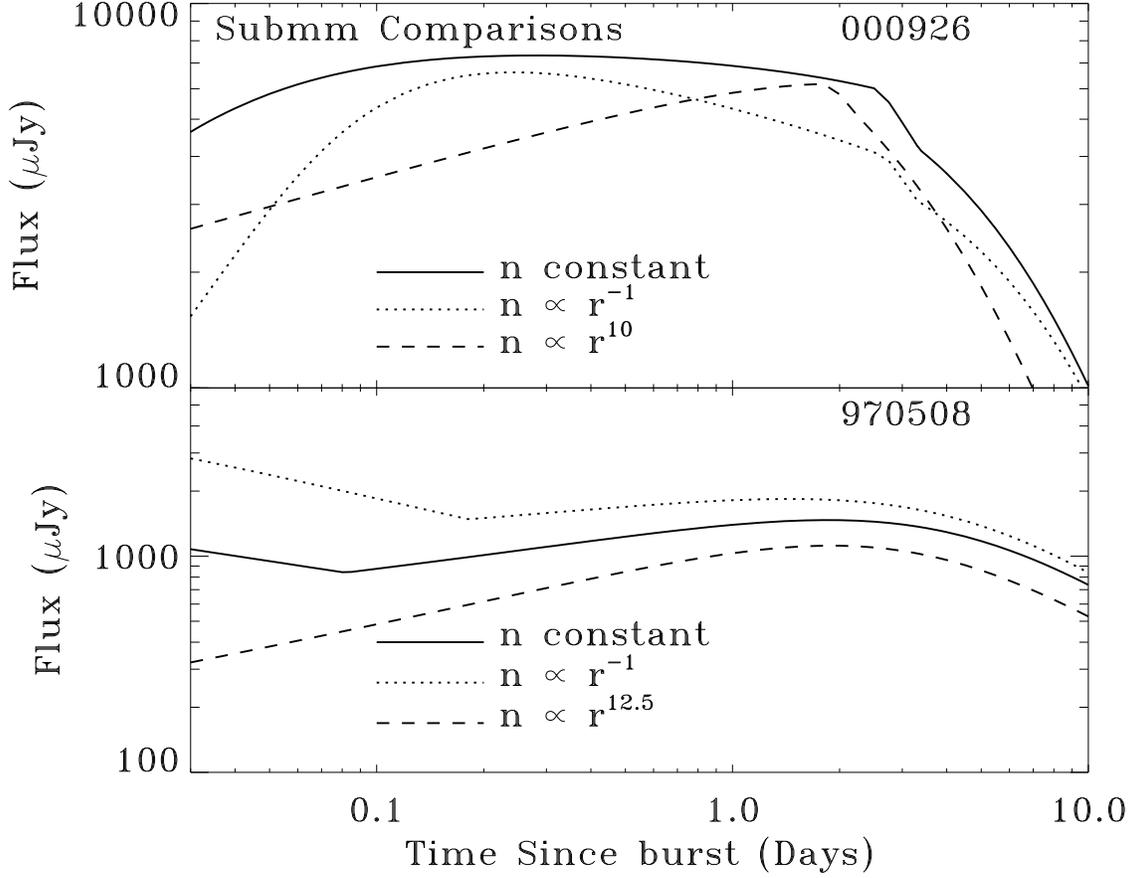}
\caption{
Comparison of model submillimeter lightcurves for several equally
acceptable model fits. The lightcurves are for a frequency of 320 GHz
(nominal center of an {\em ALMA} atmospheric window), and show the
good fits for GRBs 970508 and 000926 with the basic model as well as
with the assumption that the density $\propto r^{-1}$,$r^{\approx
10}$. Present sensitivities of $\approx$1 mJy, attainable only on
timescales $\sim$ day, are insufficient to distinguish between the
variety of peak levels that subsequently match the radio peak. This
spread in peak levels is due to differing peak behaviours (rising or
falling) whose details depend upon factors such as energy losses, jet
break, $n(r)$, or $\epsilon_B(\gamma)$. The early model divergences
due to density profile are of $\sim$ 3 mJy; with models with differing
magnetic energy fraction ($\epsilon_B(\gamma)$) it is up to 10
mJy. These early differences could be resolved with improved submm
instruments soon. The {\em ALMA} array, to be partially on-line by
2006 and completed by 2010, is expected to give fractional mJy
sensitivity in a few minutes, which could distinguish amongst these.
\label{submmcomps}}
\end{figure}

In previous sections (\S\ref{sec:epbofx}, \S\ref{sec:n_ofr}), we found
good fits to the data under different assumptions ($\epsilon_B \propto
\gamma^{-1}$, $\gamma^{+1}$; $n \propto r^{\sim 10}$, $r^{-1}$), as 
reasonable as the good fits derived from the basic model of
\S\ref{sec:basicmodel}. The datasets will tolerate widely differing 
assumptions since the fits need only match up with data over a limited
range of frequencies and times, and there are significant degeneracies
between parameters and model assumptions. For example, the decay rate
depends upon the spectral index as well as the assumed $n(r)$ or
$\epsilon_B(\gamma)$. We cross-compared these good fits with those of
the basic model, checking spectra from 0.01 to 300 days to see the
best ways of distinguishing them. The fitted models diverge in
spectral and temporal regions far from the data, as shown in Figures
\ref{xraycomps} and \ref{submmcomps}, which highlight the most
promising areas for improved constraints: longer and more sensitive
X-ray and submillimeter observations.

X-ray lightcurves of some of the acceptable fits are extended to early
times for comparison in Figure \ref{xraycomps}. Sets of acceptable
fits for two events are shown to demonstrate that the fits in all
events diverge by factors of up to three at early times; fluxes of
equally acceptable fits may be 3 $\mu$Jy or 10 $\mu$Jy at 0.01-0.03
days. The sensitivity of {\em INTEGRAL}'s instruments would not be
able to distinguish them, but more sensitive X-ray instruments on
later $\gamma$-ray missions such as {\em Swift} may. Figure
\ref{xraycomps} includes a case with an upscattered IC flux component,
whose peak passage timing gives different curvatures to the light
curves. Dense, and preferably multifrequency (the IC peak can be seen
in the spectrum), X-ray lightcurves would break the degeneracy between
synchrotron or IC flux as the X-ray source.

Observations of the broadband peak beyond the radio may be most
promising. Figure \ref{submmcomps} shows examples of the submm
lightcurves resulting from various acceptable fits found in this work;
there are a variety of peak levels that subsequently match the radio
peak level. The peak is in the mid-IR or submm for most of the
observable afterglow, from about a day to a month. It can rise or fall
for a variety of reasons (energy losses, jet break, $n(r)$,
$\epsilon_B(\gamma)$), with details dependent upon the
assumptions. These lead to models diverging in submm peak height by up
to 10 mJy at fractions of a day ($\approx$~mJy in the
mid-IR). Improved submm instruments are expected to reach appropriate
sensitivities soon. The {\em ALMA} array, to be partially on-line by
2006 and completed by 2010, is expected to give fractional mJy
sensitivity in a few minutes. Its observations could seriously
constrain the peak's behaviour.

More NIR observations will be of use; in \S\ref{sec:epbofx}
we noted that some fits break down where host extinction and $p$ could
no longer produce appropriate spectral indices both in the optical and
from the optical to the X-ray. With NIR data, host extinction is
better constrained, spectral requirements constrain $p$ and allowed
model assumptions can be better distinguished by their temporal
behaviours for that $p$.

Finally, earlier optical observations are becoming available now
(e.g. GRB~021004 \citet{fox02}, GRB~021211 \citet{fp02}) and will be
of some use. However, at such early times the dominant optical
emission should not be due to the synchrotron emission from a forward
shock into the external medium; reverse shocks (as likely seen in the
990123 optical flash, \citet{sp99b, mr99}) and internal shocks may
produce the early optical flux. This will allow further constraints,
but not necessarily to the same parameters. We may be seeing in
GRB~021004 that the forward shock dominates only after $\approx$ 0.1
days, and the rise of its peak may be masked by the reverse shock
\citep{kz03, uki+03}.

Thus, the behaviours exhibited by good fits (under various model
assumptions) to the datasets to date may be distinguishable in the
near future, especially by densely sampled X-ray lightcurves, and
observations of the peak at frequencies above the radio.

\section{Conclusions}

We fit four well-studied bursts with extensive radio through X-ray
afterglow datasets to a fireball model with simple assumptions
concerning the microphysics and environment. We find a range of
reasonable environmental and geometrical parameters. We find all four
fit best with a constant density medium, one with a value similar to
the Milky Way's ISM density, $n \approx 0.2$cm$^{-3}$, the other three
typical of diffuse clouds $n \approx 20$cm$^{-3}$. Their kinetic
energies are comparable to the total GRB $\gamma$-ray energy. The
collimation varies from near-isotropy to a half-angle of 0.04 radians.

We also find a striking diversity in the fitted microphysical
parameter values, far beyond the statistical uncertainties. The
electron energy distribution index varies from $p = 2.1 - 2.9$ and the
magnetic energy fraction varies from 0.2\% to 25\%. As shock physics
should depend merely on shock strength, we investigated whether the
spread could be due to model uncertainty, but did not find a set of
assumptions which fit the data via universal microphysics parameters.

We allowed for changes to be
made to the model assumptions: $\epsilon_B \propto \gamma^x$ and
independently $n \propto r^S$. We find considerable flexibility in the
values of $x$ and $S$ that can still produce reasonable fits:
$\epsilon_B \propto \gamma^{x}$; $-2 \leq x \leq +1$
and $n \propto r^{S}$, with $S>-2$ through $S\gg 1$. Moreover, some
parameter values change by up to an order of magnitude when the
assumptions underlying the model are altered. Clearly, even the
results of very good fits are not unique and the parameters cannot be
taken at face value. The model assumptions are not strongly
constrained by the datasets available to date.  With this model
uncertainty, the evidence for massive stellar progenitors from other
sources \citep[positions within hosts, possible SN associations,
see][]{m02} is not hard to reconcile with the lack of clear $n \propto
r^{-2}$ wind signatures in the best fits. Massive stars may not
produce a true $r^{-2}$ profile, or its effect upon the spectrum could
be masked by an inaccuracy in other model assumptions.

Finally, we compared the spectral evolution of the range of acceptable
fits with differing assumptions to identify observational strategies
that would produce better constraints. Two areas are most
promising. First, as for now a good fit need only line up with a small
time range of X-ray observations, the {\em Swift} satellite's expected
early, well-sampled X-ray lightcurves will better constrain the
spectral evolution (as well as the IC upscatters of photons to the
X-ray band and their consistency with the synchrotron model). As well,
the peak has only been definitively observed at radio frequencies,
passing through the mid-IR and submm during most of the afterglow. New
submm instruments such as {\em ALMA} should increase the reach of
direct peak detections. This will constrain the peak flux evolution,
which is sensitive to the model assumptions.

In the future it would be useful to investigate further constraints
upon the assumptions. These might include variable energy $E(t)$
\citep[under investigation in the context of events such as
021004, e.g.,][]{hp03}, and the possibility that the electron energy
(parameterized by $\epsilon_e$), or their acceleration (parameterized
by the powerlaw index $p$), could vary with shock strength.

\acknowledgements 
We thank Roger Blandford for helpful discussions.  This work was
supported in part by a NASA ATP grant awarded to RS.  FAH acknowledges
support from a Presidential Early Career Award. The National Radio
Astronomy Observatory is a facility of the National Science Foundation
operated under cooperative agreement by Associated Universities,
Inc. We made extensive use of the GCN archive, maintained by Scott
Barthelmy and the Laboratory for High-Energy Astrophysics.

\newpage
\begin{deluxetable}{ccccccccccccc}
\footnotesize
\rotate
\tabletypesize{\footnotesize}
\tablecolumns{13}
\tablewidth{0pc}
\tablecaption{Fit parameters of the best models with simple assumptions\label{tab:bestfits}}
\tablehead{
\colhead{$\chi^2$} & 
\colhead{DOF} &
\colhead{$t_{cm}$\tablenotemark{a}} &
\colhead{$t_{jet}$} &
\colhead{$t_{NR}$} &
\colhead{$E$\tablenotemark{b}} &
\colhead{$E$} &
\colhead{$\theta_{jet}$} &
\colhead{$n$} &
\colhead{A(V)} &
\colhead{$p$} &
\colhead{$\epsilon_e$} &
\colhead{$\epsilon_{B}$} \\
\colhead{} &
\colhead{} &
\colhead{} &
\colhead{} &
\colhead{} &
\colhead{$t_{cm}$} &
\colhead{1 day} &
\colhead{rad} &
\colhead{cm$^{-3}$} &
\colhead{host} &
\colhead{} &
\colhead{} &
\colhead{\%}
}
\startdata
\cutinhead{GRB 970508}
596 & 257 & 0.082& 183& 203& 3.7$^{+0.1}_{-0.1}$ & 1.6 & 0.84$^{+0.03}_{-0.03}$ & 0.20$^{+0.01}_{-0.02}$ & 0.14$^{+0.02}_{-0.02}$ & 2.1223$^{+0.003}_{-0.0008}$ & 0.342$^{+0.008}_{-0.01}$ & 25.0$^{+0.6}_{-2}$ \\
\cutinhead{GRB 980329}
 115& 90  & 6.1 &  0.12& 70 & 126$^{+6}_{-6}$ & 170 & 0.036$^{+0.002}_{-0.004}$ & 20$^{+5}_{-5}$ & 1.9$^{+0.2}_{-0.1}$ & 2.88$^{+0.1}_{-0.2}$ & 0.12$^{+0.02}_{-0.02}$ & 17$^{+3}_{-3}$  \\
\cutinhead{GRB 980703}
170 & 147   &1.4  &3.4 & 50  &11.8$^{+0.8}_{-2}$  & 13  & 0.234$^{+0.02}_{-0.007}$ & 28$^{+4}_{-3}$ & 1.15$^{+0.08}_{-0.06}$ & 2.54$^{+0.04}_{-0.1}$ & 0.27$^{+0.03}_{-0.03}$  & 0.18$^{+0.04}_{-0.03}$ \\
\cutinhead{GRB 000926}
138  &93  &3.4  &2.6  &79  &12$^{+2}_{-2}$ & 15  & 0.162$^{+0.007}_{-0.004}$ & 16$^{+3}_{-3}$ & 0.022$^{(< 0.037)}\tablenotemark{c}$ &2.79$^{+0.05}_{-0.04}$  & 0.15$^{+0.01}_{-0.01}$ &2.2$^{+0.5}_{-0.6}$  \\
\enddata
\tablenotetext{a}{Time when fast cooling ends at $\nu_c = \nu_m$}
\tablenotetext{b}{Isotropic equivalent blastwave energy (not corrected for
collimation), at the time when $\nu_c = \nu_m$. All tabled energies are in units of $10^{52}$ ergs, and isotropic-equivalent}
\tablenotetext{c}{no lower constraint on this extinction value; 68.3\% confidence interval is $<$ 0.037}
\tablecomments{Statistical uncertainties are given for the primary (employed in the fit) parameters; the other columns are derived from the fitted values. The quoted uncertainties are produced via the Monte Carlo bootstrap method with 1000 trials to generate the parameter distribution. The values bracket the resulting 68.3\% confidence interval. These error bars do not include uncertainties in the model itself. The model uncertainties are larger than the statistical uncertainties, as demonstrated from the range of parameters that produce reasonable fits under various assumptions.}
\end{deluxetable}

\newpage
\clearpage
\newpage
\begin{deluxetable}{cccc}
\footnotesize
\tablecolumns{4}
\tablewidth{0pc}
\tablecaption{Model Flux Dependences With $\epsilon_B \sim \gamma^x$\label{tab:epbx_fnu}}
\tablehead{
\colhead{Spectral Region} & 
\colhead{Parameters} & 
\colhead{$t$, $t < t_{jet}$\tablenotemark{a}} &
\colhead{$t$, $t > t_{jet}$\tablenotemark{a}} 
}
\startdata
\cutinhead{For $\nu_a~<~\nu_m~<~\nu_c$}
$\nu < \nu_a$ & $E^{0.5}n^{-0.5}\epsilon_e\epsilon_B^0$	&$t^{0.5}$ $t^{0x}$	&$t^0$ $t^{0x}$ \\
$\nu_a < \nu < \nu_m$ & $E^{0.83}n^{0.5}\epsilon_e^{-0.67}\epsilon_B^{0.33}$	&$t^{0.5}$ $t^{-0.13x}$	&$t^{-0.33}$ $t^{-0.17x}$ \\
$\nu_m < \nu < \nu_c$ & $E^{1.35}n^{0.5}\epsilon_e^{1.4}\epsilon_B^{0.85}$	& $t^{-1.05}$ $t^{-0.32x}$	&$t^{-2.4}$ $t^{-0.43x}$ \\
$\nu_c < \nu$\tablenotemark{b} & $E^{1.1}n^0\epsilon_e^{1.4}\epsilon_B^{0.1}$	&$t^{-1.3}$ $t^{-0.04x}$	&$t^{-2.4}$ $t^{-0.05x}$ \\
\cutinhead{For $\nu_a~<~\nu_c~<~\nu_m$}
$\nu < \nu_a$\tablenotemark{b} & $E^{0}n^{-1}\epsilon_e^{0}\epsilon_B^{-1}$	&$t^{1}$ $t^{0.38x}$	&$t^{1}$ $t^{0.5x}$ \\
$\nu_a < \nu < \nu_c$\tablenotemark{b} & $E^{1.17}n^{0.83}\epsilon_e^0\epsilon_B$	&$t^{0.17}$ $t^{-0.38x}$	&$t^{-1}$ $t^{-0.5x}$ \\
$\nu_c < \nu < \nu_m$\tablenotemark{b} & $E^{0.75}n^0\epsilon_e^0\epsilon_B^{-0.25}$	&$t^{-0.25}$ $t^{0.09x}$	&$t^{-1}$ $t^{0.13x}$ \\
$\nu_m < \nu$\tablenotemark{b} & $E^{1.1}n^0\epsilon_e^{1.4}\epsilon_B^{0.1}$	&$t^{-1.3}$ $t^{-0.04x}$	&$t^{-2.4}$ $t^{-0.05x}$ \\
\enddata
\tablenotetext{a}{p=2.4 is assumed where necessary}
\tablenotetext{b}{for synchrotron cooling dominating; $\nu_c$
behaviour changes for IC-dominant cooling, see section \ref{sec:emission} for details}
\end{deluxetable}

\newpage
\clearpage
\begin{deluxetable}{cccccccccccccc}
\rotate
\tabletypesize{\footnotesize}
\tablecolumns{14}
\tablewidth{0pc}
\tablecaption{Fit parameters for the best models with $\epsilon_B = K \gamma^x$\label{tab:epbx_bestfits}}
\tablehead{
\colhead{$x$} & 
\colhead{$\chi^2$} & 
\colhead{DOF} &
\colhead{$t_{cm}$\tablenotemark{a}} &
\colhead{$t_{jet}$} &
\colhead{$t_{NR}$} &
\colhead{$E$\tablenotemark{b}} &
\colhead{$E$} &
\colhead{$n$} &
\colhead{$p$} &
\colhead{$\epsilon_{B,\%}$} &
\colhead{$K$=$\epsilon_{B,\%}$/$\gamma^x$} &
\colhead{$\epsilon_e$} &
\colhead{$\theta$\tablenotemark{c}}\\
\colhead{} &
\colhead{} &
\colhead{} &
\colhead{(d)} &
\colhead{(d)} &
\colhead{(d)} &
\colhead{$t_{cm}$} &
\colhead{1 day} &
\colhead{cm$^{-3}$} &
\colhead{} &
\colhead{1 day} &
\colhead{} &
\colhead{} &
\colhead{}
}
\startdata
\cutinhead{GRB 970508}
$+2$ & 945 & 257 & 1.7 & 223& 223& 1.6& 1.8& 0.87& 2.23 & MAX & 2.2& 0.19& ISO\\
$+1$ & 695 & 257 & 0.57 & 322& 322& 1.8& 1.5& 0.21& 2.20 & 80 & 8.2& 0.24& ISO\\
 $0$  & 596 & 257 & 0.082& 183& 203& 3.7& 1.6& 0.20& 2.12 & 25 &25 & 0.34& 0.84\\
$-1$ & 600 & 257 & 0.028& 246 & 246& 7.1& 1.6& 0.15& 2.09 & 22 & 220  & 0.45& ISO\\
-2\tablenotemark{d} & 569 & 257 & 0.0015 & 289 & 289 & 27 & 1.5 & 0.10 & 2.07 & 21 & 2400 & 0.59 & ISO\\
-2\tablenotemark{e} & 703 & 257 & 0.0011 & 498 & 498 & 13 & 2.0 & 0.032 & 2.14 &  MAX & 5.9e4 & 0.24 & ISO\\
\cutinhead{GRB 980329}
+4 & 112 & 90 & 6.5 & 76 & 76 & 2.9 & 4.7  & 220 & 2.31  & 21 & 0.021 & 0.23 & ISO\\
 +1 & 117 & 90 & 0.30 & 1.1 &151 &4100&3900& 150 & 2.0007 & 64 & 4.7  &0.027  &0.075 \\
0 &  115& 90  & 6.1 &  0.12& 70 & 130 &170& 20 & 2.88   & 17 & 17 & 0.12 & 0.036\\
-1 & 107 &90  & 0.66 & 0.16 & 79 & 174 & 170 & 15 & 2.56 & 3.0 & 29 &0.061  & 0.040\\
-4\tablenotemark{d} & 114 & 90 & - & 0.38 & 73 & 50 & 27 & 6.4 & 2.35  & MAX & 2.7e13 & 0.090 & 0.064\\
-4\tablenotemark{e} & 124 & 90 & - & 0.45 & 83 & 2000 & 35 & 6.9 & 2.43  & MAX & 1.2e13 & 0.074 & 0.066\\
\cutinhead{GRB 980703}
+3 & 230 & 147 & 3.4 & 121 & 121 & 0.74 & 1.2 & 2.1 & 2.13 & MAX & 1.3 & 0.39 &ISO \\
+1 &194  & 147 &1.3  &75  &85  &1.4 &1.6 &1.2 &2.05  & MAX & 13  &0.68 &0.83 \\
0 & 170 & 147   &1.4  &3.4 &50  &12  &13  &28  &2.54 & 0.18 & 0.18 &0.27 &0.23 \\
-1 & 165 & 147 & 1.2 & 3.5 & 52 & 14 & 16 & 22 & 2.21 & 0.083 & 0.62 & 0.51 & 0.23\\
-3\tablenotemark{d} & 180 & 147 & 0.026 & 1.7 & 76 & 120 & 12 & 2.3 & 2.06 &  0.42 & 370 & 1 & 0.13\\
-3\tablenotemark{e} & 174 & 147 & 0.044 & 1.9 & 76 & 170 & 20 & 3.8 & 2.06 &  0.16 & 140 & 1 & 0.14\\
-(4$^{+}$) & 180 & 147 & 0.36 & 73 & 93 & 3.8 & 1.9 & 0.51 & 2.03 & MAX & 1.5e6 & 0.92 & 0.78\\
\cutinhead{GRB 000926}
+1 &209  &93 &6.0  &8.2  &80  &3.8 &5.3 &14  &2.88  & 20 & 2.5  &0.15  &0.28 \\
0 & 138  &93  &3.4  &2.6  &79  &12 &15  &16  &2.79  & 2.2 & 2.2  &0.15  &0.16 \\
-1 & 127  &93  &3.4  &1.7  &80  &23 &32  &23  &2.61 & 0.26 & 2.5 &0.23  &0.13 \\
-3\tablenotemark{d} & 198 & 93 & 0.043 & 0.79 & 100 & 28 & 15 & 2.4 & 2.18 & 11 & 1.5e4 & 0.20 & 0.079\\
-3\tablenotemark{e} & 217 & 93 & 0.25 & 0.25 & 111 & 85 & 66 & 2.7 & 2.21 & 6.6 & 1.0e4 & 0.15 & 0.042 \\
\enddata
\tablenotetext{a}{Time when fast cooling ends at $\nu_c$ = $\nu_m$}
\tablenotetext{b}{Isotropic equivalent blastwave energy (not corrected for
collimation), at the time when $\nu_c$ = $\nu_m$. All tabled energies in units of $10^{52}$ ergs, and isotropic-equivalent}
\tablenotetext{c}{Jet half-opening angle, in radians; $>$ 1 radian treated as isotropic}
\tablenotetext{d}{$x=0$ quenching rate of energy losses}
\tablenotetext{e}{No E(t) quenching, see \S\ref{sec:epbofx} for details}
\tablecomments{Certain items reach limits indicated in the table. Collimations reaching the isotropic limit of $\theta \approx 1$ are noted as ``ISO''. Some magnetic energy fractions reach the physical limit of 100\% of the shock energy, noted as ``MAX''; for such models where the magnetic energy fraction drops with the shock $\gamma$ ($x>0$), $\epsilon_B$ is in the physical range over much of the data's time range, the other such models are pinned at this limit over the entire dataset. Blank times $t_{cm}$ for the end of fast cooling indicate that the afterglow model formally has the transition much earlier than any afterglow data, in the range of prompt emission.}
\end{deluxetable}

\newpage
\begin{deluxetable}{ccccccccccccccc}
\footnotesize
\rotate
\tabletypesize{\scriptsize}
\tablecolumns{15}
\tablewidth{0pc}
\tablecaption{Best fits assuming $\epsilon_B \propto \gamma^{-1}$\label{tab:bestxn1}}
\tablehead{
\colhead{$\chi^2$} & 
\colhead{DOF} &
\colhead{$t_{cm}$\tablenotemark{a}} &
\colhead{$t_{jet}$} &
\colhead{$t_{NR}$} &
\colhead{$E_{52}$\tablenotemark{b}} &
\colhead{$E_{52}$} &
\colhead{$\theta_{jet}$} &
\colhead{$n$} &
\colhead{A(V)} &
\colhead{$p$} &
\colhead{$\epsilon_e$} &
\colhead{$\epsilon_{B,\%}$} &
\colhead{$\epsilon_{B,\%}$} &
\colhead{$\epsilon_{B,\%}$} \\
\colhead{} &
\colhead{} &
\colhead{(d)} &
\colhead{(d)} &
\colhead{(d)} &
\colhead{$t_{cm}$} &
\colhead{1 d} &
\colhead{rad} &
\colhead{cm$^{-3}$} &
\colhead{host} &
\colhead{} &
\colhead{} &
\colhead{$t_{cm}$} &
\colhead{1 d} &
\colhead{$t_{NR}$}
}
\startdata
\cutinhead{GRB 970508}
600 & 257 & 0.028 & 246 & 246 & 7.1$^{+1.2}_{-0.3}$ & 1.6 & ISO\tablenotemark{c}  &  0.146$^{+0.004}_{-0.04}$ & 0.09$^{+0.03}_{-0.04}$  & 2.088$^{+0.006}_{-0.002}$ & 0.45$^{+0.02}_{-0.03}$  & 4.8$^{+0.4}_{-1.1}$ & 22 & 100 \\
\cutinhead{GRB 980329}
107 & 90 & 0.66 & 0.16 & 79 & 174$^{+6}_{-11}$  & 170 & 0.040$^{+0.004}_{-0.002}$  & 15$^{+4}_{-2}$  & 1.46$^{+0.09}_{-0.08}$  & 2.56$^{+0.05}_{-0.08}$  & 0.061$^{+0.008}_{-0.01}$  & 2.4$^{+0.5}_{-1.1}$  & 3.0 & 29 \\
\cutinhead{GRB 980703}
165 & 147 & 1.2 & 3.5 & 52 & 14$^{+2}_{-1}$  & 16 & 0.23$^{+0.02}_{-0.01}$ & 22$^{+4}_{-3}$ & 1.25$^{+0.05}_{-0.05}$  & 2.21$^{+0.04}_{-0.03}$ & 0.51$^{+0.04}_{-0.06}$ & 0.091$^{+0.010}_{-0.018}$ & 0.083 & 0.62 \\
\cutinhead{GRB 000926}
127 & 93 & 3.4 & 1.7 & 80 & 23$^{+2}_{-3}$ & 32 & 0.131$^{+0.004}_{-0.005}$  & 23$^{+4}_{-5}$ & 0.036$^{(<0.063)}$\tablenotemark{d} & 2.61$^{+0.08}_{-0.2}$ & 0.23$^{+0.05}_{-0.03}$ & 0.47$^{+0.14}_{-0.15}$ & 0.26 & 2.5 \\
\enddata
\tablenotetext{a}{Time when fast cooling ends at $\nu_c = \nu_m$}
\tablenotetext{b}{Isotropic equivalent blastwave energy (not corrected for
collimation), at the time when $\nu_c = \nu_m$. All tabled energies are in units of $10^{52}$ ergs, and isotropic-equivalent}
\tablenotetext{c}{No results with jet half-opening angles $<$ 1 radian; all treated as isotropic}
\tablenotetext{d}{no lower constraint on this extinction value; 68.3\% confidence interval is $<$ 0.063}
\tablecomments{Statistical uncertainties are given for the primary (employed in the fit) parameters; the other columns are derived from the fitted values. The quoted uncertainties are produced via the Monte Carlo bootstrap method with 1000 trials to generate the parameter distribution. The values bracket the resulting 68.3\% confidence interval. These error bars do not include uncertainties in the model itself. The model uncertainties are larger than the statistical uncertainties, as demonstrated from the range of parameters that produce reasonable fits under various assumptions.}
\end{deluxetable}

\newpage
\begin{deluxetable}{ccccccccccccccccc}
\rotate
\footnotesize
\tabletypesize{\footnotesize}
\tablecolumns{17}
\tablewidth{0pc}
\tablecaption{Fit parameters for the best models with  $n = n_i(r/r_i)^S$\label{tab:nofr_fits}}
\tablehead{
\colhead{$S$} & 
\colhead{$\chi^2$} & 
\colhead{DOF} &
\colhead{$t_{cm}$\tablenotemark{a}} &
\colhead{$t_{jet}$} &
\colhead{$t_{NR}$} &
\colhead{$E_{cm}$\tablenotemark{b}} &
\colhead{$E$} &
\colhead{$n_{18}$\tablenotemark{c}} &
\colhead{$n$\tablenotemark{d}} &
\colhead{$R_{18}$\tablenotemark{e}} &
\colhead{$n$\tablenotemark{d}} &
\colhead{$R_{18}$\tablenotemark{e}} &
\colhead{$p$} &
\colhead{$\epsilon_{B}$} &
\colhead{$\epsilon_e$} &
\colhead{$\theta$\tablenotemark{f}}\\
\colhead{} &
\colhead{} &
\colhead{} &
\colhead{(d)} &
\colhead{(d)} &
\colhead{(d)} &
\colhead{} &
\colhead{1 d} &
\colhead{} &
\colhead{1 d} &
\colhead{1 d} &
\colhead{100 d} &
\colhead{100 d} &
\colhead{} &
\colhead{(\%)} &
\colhead{} &
\colhead{}
}
\startdata
\cutinhead{GRB 970508}
12.5 & 617 & 257 & 0.0044 & 52 & 52 & 8.3 & 0.75 & 0.087 & 0.14 & 1.0 & 1.9 & 1.3 & 2.08 & 3.2  & 0.56 & ISO\\
0  & 596 & 257 & 0.082& 183& 203& 3.7& 1.6 & 0.20 & 0.20 & 0.57 & 0.20 & 2.7 & 2.12& 25 & 0.34& 0.84\\
-2.5 & 832 & 257 & 4.5 & $1.3\times 10^{5}$ & $1.3\times 10^{5}$ & 0.44 & 0.50 & 0.021 & 3.3 & 0.13 & 0.0023 & 2.4 & 2.32 & MAX & 0.052 & ISO\\
-2.5 & 1012 & 257 & 3.8 & $6.2\times 10^{3}$ & $6.2\times 10^{3}$ & 1.8 & 2.6 & 0.25 & 170 & 0.0074 & 0.30 & 0.93 & 2.28 & 0.095 & 0.20 & ISO\\
\cutinhead{GRB 000926}
12 & 160 & 93 & 1.8 & 1.6 & 23 & 4.3 & 5.4 & $9.4\times 10^{7}$ & 70 & 0.31 & 320 & 0.35 & 2.14 & 0.34 & 0.35 & 0.27\\
12 & 141 & 93 & 2.2 & 2.2 & 26 & 7.5 & 9.7 & $1.9\times 10^{7}$ & 73 & 0.35 & 530 & 0.42 & 2.64 & 0.026 & 0.31 & 0.29\\
10 & 135 & 93 & 2.1 & 1.9 & 26 & 6.7 & 8.2 & $1.6\times 10^{7}$ & 92 & 0.30 & 560 & 0.36 & 2.45 & 0.046 & 0.26 & 0.27\\
0 & 138  &93  &3.4  &2.6  &79  &12 &15  & 16 & 16 & 0.29 & 16 & 0.37 & 2.79  & 2.2  &0.15  &0.16 \\
-2.5 & 196 & 93 & 0.35 & $4.7\times 10^{7}$ & $4.7\times 10^{7}$ & 48 & 47 & 0.17 & 1.2 & 0.46 & 0.00056 & 9.9 & 2.88 & 0.046 & 0.033 & ISO\\
\enddata
\tablenotetext{a}{Time when fast cooling ends at $\nu_c = \nu_m$}
\tablenotetext{b}{Isotropic equivalent blastwave energy (not corrected for
collimation), at the time when $\nu_c = \nu_m$. All tabled energies are in units of $10^{52}$ ergs, and isotropic-equivalent}
\tablenotetext{c}{Density at $R = 10^{18}$cm, in units of cm$^{-3}$, the fit parameter}
\tablenotetext{d}{in units of cm$^{-3}$}
\tablenotetext{e}{Radius in units of 10$^{18}$ cm, as calculated, not a fit parameter}
\tablenotetext{f}{Jet half-opening angle, in radians}
\tablecomments{Certain items reach limits indicated in the table. Collimations reaching the isotropic limit of $\theta \approx 1$ are noted as ``ISO''. In one model the magnetic energy fraction reaches the physical limit of 100\% of the shock energy, noted as ``MAX''. The densities given at 1 and 100 days give an idea of the range of density probed in the model. These are calculated post-fit; the fit uses scalings based upon the density powerlaw, not a density calculation at each time. }
\end{deluxetable}

\newpage
\clearpage
\newpage
\begin{deluxetable}{cccc}
\footnotesize
\tablecolumns{4}
\tablewidth{0pc}
\tablecaption{Model Flux Dependences With $n = n_i(r/r_i)^S$\label{tab:fnu_nofr}}
\tablehead{
\colhead{Spectral Region} & 
\colhead{Parameters} & 
\colhead{$t$, $t < t_{jet}$\tablenotemark{a}} &
\colhead{$t$, $t > t_{NR}$\tablenotemark{a}} 
}
\startdata
\cutinhead{For $\nu_a~<~\nu_m~<~\nu_c$}
$\nu < \nu_a$ & $E^{\frac{2}{4+s}}n_i^{\frac{-2}{4+s}}\epsilon_e\epsilon_B^0$	&$t^{\frac{2}{4+s}}$ & $t^{\frac{-2(s+1)}{s+5}}$\\
$\nu_a < \nu < \nu_m$ & $E^{\frac{10+4s}{12+3s}}n_i^{\frac{2}{4+s}}\epsilon_e^{\frac{-2}{3}}\epsilon_B^{\frac{1}{3}}$	&$t^{\frac{s+2}{s+4}}$	&$t^{\frac{10s+24}{3s+15}}$ \\
$\nu_m < \nu < \nu_c$ & $E^{\frac{p}{4}+\frac{12+5s}{16+4s}}n_i^{\frac{2}{s+4}}\epsilon_e^{p-1}\epsilon_B^{\frac{p+1}{4}}$	& $t^{\frac{-3p}{4} + \frac{12+5s}{16+4s}}$ &$t^{\frac{8s+21-(4s+15)p}{2(s+5)}}$ \\
$\nu_c < \nu$\tablenotemark{b} & $E^{\frac{p+2}{4}}n_i^0\epsilon_e^{p-1}\epsilon_B^{\frac{p-2}{4}}$	&$t^{\frac{-3p+2}{4}}$ &$t^{\frac{6s+20-(4s+15)p}{2(s+5)}}$ \\
\cutinhead{For $\nu_a~<~\nu_c~<~\nu_m$}
$\nu < \nu_a$\tablenotemark{b} & $E^{\frac{-s}{4+s}}n_i^{\frac{-12}{12+3s}}\epsilon_e^{0}\epsilon_B^{-1}$	&$t^{\frac{4}{4+s}}$ &$t^{\frac{8s+10}{3s+15}}$ \\
$\nu_a < \nu < \nu_c$\tablenotemark{b} & $E^{\frac{14+6s}{12+3s}}n_i^{\frac{10}{12+3s}}\epsilon_e^0\epsilon_B$	&$t^{\frac{2+3s}{12+3s}}$	&$t^{\frac{2s+5}{2s+10}}$ \\
$\nu_c < \nu < \nu_m$\tablenotemark{b} & $E^{\frac{3}{4}}n_i^0\epsilon_e^0\epsilon_B^{\frac{-1}{4}}$	&$t^{\frac{1}{4}}$ & $t^{\frac{-s+5}{s+5}}$ \\
$\nu_m < \nu$\tablenotemark{b} & $E^{\frac{p+2}{4}}n_i^0\epsilon_e^{p-1}\epsilon_B^{\frac{p-2}{4}}$	&$t^{\frac{-3p+2}{4}}$ & $t^{\frac{6s+20-(4s+15)p}{2(s+5)}}$\\
\enddata
\tablenotetext{a}{to the model approximation, $r$ is constant during the jet spreading phase, and so the behaviour at $t_{jet}<t<t_{NR}$ is the same for any density profile (see table \ref{tab:epbx_fnu}, for $x=0$)}
\tablenotetext{b}{for synchrotron cooling dominating; $\nu_c$
behaviour changes for IC-dominant cooling, see section \ref{sec:emission} for details}
\end{deluxetable}

\newpage
\clearpage
\newpage

\newpage
\clearpage
\newpage

\newpage
\clearpage
\newpage

\newpage
\clearpage
\newpage

\newpage
\clearpage
\newpage

\newpage
\clearpage
\newpage

\newpage
\clearpage
\newpage

\newpage
\clearpage
\newpage

\end{document}